\documentclass[twocolumn]{aastex6}

\usepackage{multirow}
\usepackage{amssymb,amsmath}
\usepackage{graphicx}
\usepackage{subfigure}

\newcommand{\hst}{\textit{HST}}
\newcommand{\Msun}{M_{\odot}}
\newcommand{\kms}{\>{\rm km}\,{\rm s}^{-1}}
\newcommand{\masyr}{\>{\rm mas}\,{\rm yr}^{-1}}
\newcommand{\musyr}{\>{\mu \rm as}\,{\rm yr}^{-1}}
\newcommand{\muw}{\mu_{W}}
\newcommand{\mun}{\mu_{N}}
\newcommand{\kpc}{\>{\rm kpc}}
\renewcommand{\vec}[1]{\mathbf{#1}}

\shorttitle{HST Proper Motions of the Draco and Sculptor dSphs}
\shortauthors{Sohn et al.}

\begin{document}

\title{Space Motions of the Dwarf Spheroidal Galaxies Draco and Sculptor \\
       based on {\it HST} Proper Motions with $\sim10$-year Time Baseline}

\author{
Sangmo Tony Sohn\altaffilmark{1}, 
Ekta Patel\altaffilmark{2}, 
Gurtina Besla\altaffilmark{2}, 
Roeland P. van der Marel\altaffilmark{1,3},
James S. Bullock\altaffilmark{4},\\
Louis E. Strigari\altaffilmark{5},
Glenn van de Ven\altaffilmark{6},
Matt G. Walker\altaffilmark{7},
\&
Andrea Bellini\altaffilmark{1}
}

\altaffiltext{1}{Space Telescope Science Institute, 
                 3700 San Martin Drive, Baltimore, MD 21218, USA}
\altaffiltext{2}{Department of Astronomy, The University of Arizona,
                 933 N. Cherry Avenue, Tucson, AZ 85721, USA}
\altaffiltext{3}{Center for Astrophysical Sciences, Department of Physics and Astronomy,
                 Johns Hopkins University,
                 Baltimore, MD 21218, USA}
\altaffiltext{4}{Center for Cosmology, Department of Physics and Astronomy, 
                 University of California,
                 Irvine, CA 92697, USA}
\altaffiltext{5}{Department of Physics and Astronomy, 
                 Mitchell Institute for Fundamental Physics and Astronomy, 
                 Texas A\&M University,
                 College Station, TX 77843-4242, USA}
\altaffiltext{6}{Max Planck Institute for Astronomy,
                 K\"{o}nigstuhl 17, D-69117 Heidelberg, Germany}
\altaffiltext{7}{McWilliams Center for Cosmology, Department of Physics,
                 Carnegie Mellon University,
                 5000 Forbes Avenue, Pittsburgh, PA 15213, USA}

\begin{abstract}
We present new proper motion (PM) measurements of the dwarf spheroidal 
galaxies (dSphs) Draco and Sculptor using multi-epoch images obtained with 
the {\it Hubble Space Telescope} ACS/WFC. Our PM results have uncertainties 
far lower than previous measurements, even made with the same instrument. 
The PM results for Draco and Sculptor are $(\mu_{W},\>\mu_{N})_{\rm Dra} 
= (-0.0562 \pm 0.0099,\>-0.1765 \pm 0.0100)\ {\rm mas\ yr}^{-1}$ and 
$(\mu_{W},\>\mu_{N})_{\rm Scl} = (-0.0296 \pm 0.0209,\>-0.1358 \pm 0.0214)\ 
{\rm mas\ yr}^{-1}$. The implied Galactocentric velocity vectors for Draco 
and Sculptor have radial and tangential components: 
$(V_{\rm rad},\>V_{\rm tan})_{\rm Dra} = (-88.6,\>161.4) \pm (4.4,\>5.6) \kms$; 
and $(V_{\rm rad},\>V_{\rm tan})_{\rm Scl} = (72.6,\>200.2) \pm (1.3,\>10.8) \kms$.
We study the detailed orbital histories of both Draco and Sculptor via 
numerical orbit integrations. Orbital periods of Draco and Sculptor are found 
to be 1--2 and 2--5 Gyrs, respectively, accounting for uncertainties in the 
MW mass. We also study the influence of the Large Magellanic Cloud (LMC) on 
the orbits of Draco and Sculptor. Overall, the inclusion of the LMC increases 
the scatter in the orbital results. Based on our calculations, Draco shows a 
rather wide range of orbital parameters depending on the MW mass and 
inclusion/exclusion of the LMC, but Sculptor's orbit is very well constrained 
with its most recent pericentric approach to the MW being 0.3--0.4 Gyr ago. 
Our new PMs imply that the orbital trajectories of both Draco and Sculptor 
are confined within the Disk of Satellites (DoS), better so than implied by 
earlier PM measurements, and likely rule out the possibility that these two 
galaxies were accreted together as part of a tightly bound group. 
\end{abstract}


\section{Introduction}

The orbital histories of Milky Way (MW) satellites contain crucial 
information about the formation and assembly history of the MW halo.
Direct access to proper motion (PM) measurements is required to derive 
their orbits. Despite the various efforts to measure PMs of 
MW satellites in the past decade or so, both the quantity and the 
quality of measurements are still lacking. The only clear solution to this 
problem is to directly measure PMs of tracer objects, but this has been 
technically challenging due to the difficulty in measuring very small 
apparent motions. The excellent astrometric capability of \hst\ has eased the 
situation, and combined with our PM measurement technique using background 
galaxies as stationary reference sources \citep{soh12}, we are now able to 
reach unprecedented PM accuracies using multi-epoch \hst\ data. 

As part of our HSTPROMO collaboration \citep{vdm14}, we are carrying out 
\hst\ programs to measure PMs of MW satellite objects. For example, we 
measured the PM of Leo~I using multi-epoch ACS/WFC images separated by 5 
years in time, and explored its orbits under realistic MW potentials 
\citep{soh13}. In addition, by comparing the observed parameters based on 
our PM measurements to Leo~I-like subhalos found in cosmological simulations, 
we constrained the virial mass of the MW \citep{boy13}. We are continuing 
to measure PMs of distant satellites in the MW halo, including ultra-faint 
dwarfs and classical dwarf spheroidal galaxies (dSphs). This study focuses 
on two classical dSphs, Draco and Sculptor.

Draco and Sculptor are located at distances of 76 and 86 kpc, respectively. 
As dynamical tracers, they probe the MW mass at important distances where 
there are only a limited number of tracer objects with tangential velocities.
The MW mass is generally estimated through equilibrium modeling using 
observed dynamical properties of halo tracers like satellites. 
Without the knowledge of tangential velocities, however, the mass estimates 
suffer from the (in)famous mass-anisotropy degeneracy. \citet{wat10} used 
older PM measurements of Draco and Sculptor \citep{sch94,pia06}, along with 
those of other dwarf satellites, to estimate the mass of the MW. However, 
the quality of these older PM measurements have limited their ability to 
constrain the MW mass with high confidence. 

Draco and Sculptor are interesting because, while they are found at similar 
Galactocentric distances, they appear to be orbiting around the MW in 
opposite directions. This was first noted by \citet{pry08}%, and later by \citet{met08}, 
who analyzed the space motions of galaxies with the PMs that existed 
at the time. Many of the satellite galaxies of the MW are found to be 
distributed on a ``Disk of Satellites (DoS)'', an orbital plane claimed to be 
occupied by most of the classical dwarf satellites of the MW \citep{lyn76,kro05,met07}. 
Most dwarfs that are believed to be members of the DoS and also have PM 
measurements are found to orbit in the same direction about the MW, with the 
notable exception of Sculptor. Interestingly, Sculptor seems to be orbiting 
around the plane in the {\it opposite} direction of most satellites. 
\citet{paw11} tried to explain this in the context of tidal-dwarf galaxies. 
Better PM measurements for a representative galaxy that rotates along the 
plane (Draco), and for a galaxy that seems to counter-rotate (Sculptor) are 
needed to shed new light on this matter.

This picture is further complicated by perturbations from the MW's most 
massive satellite galaxy, the Large Magellanic Cloud (LMC). It is unclear 
whether the gravitational pull of the LMC might complicate the orbits of the 
classical satellites, causing their membership to the plane of satellites to 
be unstable. The recent capture of the LMC by the MW \citep{bes07,kal13} may 
limit its dynamical influence on the DoS members, but this is impossible to 
properly assess without accurate PM measurements for the classical dwarfs.
Draco and Sculptor, with their opposite sense of motions, and the coincidence 
of their orbital planes with that of the LMC, present an ideal test case 
for the influence of the LMC on the DoS.

As with other satellite objects in the MW halo, the first PM measurements 
of Draco and Sculptor were carried out using photographic plates 
\citep{sch94,sch95}. The quality of these measurements is poor by  
modern standards.\footnote{The one-dimensional PM uncertainties were 
$0.35 \masyr$ for Draco and $0.24 \masyr$ for Sculptor.}
It was not until the use of \hst\ that PM uncertainties were small enough
that the results were meaningful. \citet{pia06} used multi-epoch imaging 
data obtained with the Space Telescope Imaging Spectrograph (STIS) onboard 
\hst\ to measure the PM of Sculptor. They used quasi-stellar objects 
(QSOs) in two fields as stationary reference sources to reach a final 
1-d PM uncertainty of $0.13 \masyr$. For Draco, \citet{pry15} measured 
the PM using both QSOs and background galaxies in a single field to 
achieve a 1-d PM uncertainty of $0.063 \masyr$. Meanwhile, \citet{cas16} 
used ground-based images obtained with the {\it Subaru} Suprime-Cam 
to measure the PM of Draco with a smaller 1-d PM uncertainty of 
$0.044 \masyr$. Notwithstanding the $\sim 6\sigma$ level discrepancy 
found against the \hst\ measurement by \citet{pry15}, this study 
demonstrates what can be achieved using wide-field detectors on an 
8-meter class telescope when extensive calibrations are carried out. 
It also provides hints on what to expect in the {\it LSST} era for 
PM measurements with large telescopes. 

In this paper, we present our new PM measurements for Draco and Sculptor 
using multi-epoch \hst\ imaging data separated by $\sim 10$ years in time. 
This paper is organized as follows. In Section~\ref{s:pm}, we describe 
the data, outline the analysis steps, and report the results of our PM 
measurements. In Section~\ref{s:spacemotions}, we derive the 
Galactocentric space motions of Draco and Sculptor by correcting the 
measured PMs for the solar motions. In Section~\ref{s:orbits}, we 
explore the implications for the past orbits of Draco and Sculptor 
under various assumptions for the mass of the MW, and also explore 
the gravitational influence of the Large Magellanic Cloud (LMC) on 
their orbits. Finally, in Section~\ref{s:conclusions}, we summarize 
the main results of our paper.

\section{Proper Motions}\label{s:pm}

\subsection{Data}\label{ss:data}
%123456789012345678901234567890123456789012345678901234567890123456789

%%%%%%%%%%%%%%%%%%%%%%%%%%%%%%%%%%%%%%%%%%%%%%%%%%%%%%%%%%%%%%%%%%%%%%
%% TABLE 1
%%%%%%%%%%%%%%%%%%%%%%%%%%%%%%%%%%%%%%%%%%%%%%%%%%%%%%%%%%%%%%%%%%%%%%
%
\begin{deluxetable*}{lcccccccc}
\tablecaption{Observation summary of the Draco and Sculptor dSphs
              \label{t:obslog}}
\tablehead{
\colhead{}      &                   &                   & \multicolumn{2}{c}{Epoch~1}                  &\colhead{} & \multicolumn{2}{c}{Epoch~2}                 \\
\cline{4-5}\cline{7-9}
\colhead{}      & \colhead{R.A.}    & \colhead{Decl.}   & \colhead{Date}    & \colhead{Exp. Time}      &\colhead{} & \colhead{Date}    & \multicolumn{2}{c}{Exp. Time} \\
\colhead{Field} & \colhead{(J2000)} & \colhead{(J2000)} & \colhead{(Y-M-D)} & \colhead{(s$\times N$)}  &\colhead{} & \colhead{(Y-M-D)} & \multicolumn{2}{c}{(s$\times N$)} 
          }
\startdata
%\sidehead{{\bf Draco}}
{\bf Draco}    &             &               &            & \bf{F606W}        & &            & \bf{F606W}     & \bf{F814W}    \\
{\it F1}       & 17:21:01.34 & $+$57:58:38.5 & 2004-10-19 & 430s$\times$19    & & 2013-10-14 & 453s$\times$12 & 300s$\times$3 \\
{\it F2}       & 17:21:51.69 & $+$58:01:41.0 & 2004-10-31 & 430s$\times$19    & & 2012-10-24 & 501s$\times$12 & 300s$\times$3 \\
{\it F3}       & 17:19:29.97 & $+$57:58:10.2 & 2004-10-30 & 430s$\times$19    & & 2012-10-26 & 507s$\times$12 & 300s$\times$3 \\
\hline
%\sidehead{{\bf Sculptor}}
{\bf Sculptor} &             &               &            & \bf{F775W}        & &            & \bf{F775W}     & \bf{F606W}    \\
{\it F1}       & 00:59:57.31 & $-$33:46:23.5 & 2002-09-28 & 417s$\times$\phn5 & & 2013-09-29 & 419s$\times$16 & 150s$\times$4 \\
{\it F2}       & 00:59:48.61 & $-$33:48:47.1 & 2002-09-26 & 400s$\times$\phn6 & & 2013-09-29 & 419s$\times$16 & 150s$\times$4 \\
\enddata
\end{deluxetable*}
%
%%%%%%%%%%%%%%%%%%%%%%%%%%%%%%%%%%%%%%%%%%%%%%%%%%%%%%%%%%%%%%%%%%%%%%

\begin{figure*}
\epsscale{1.15}
\plottwo{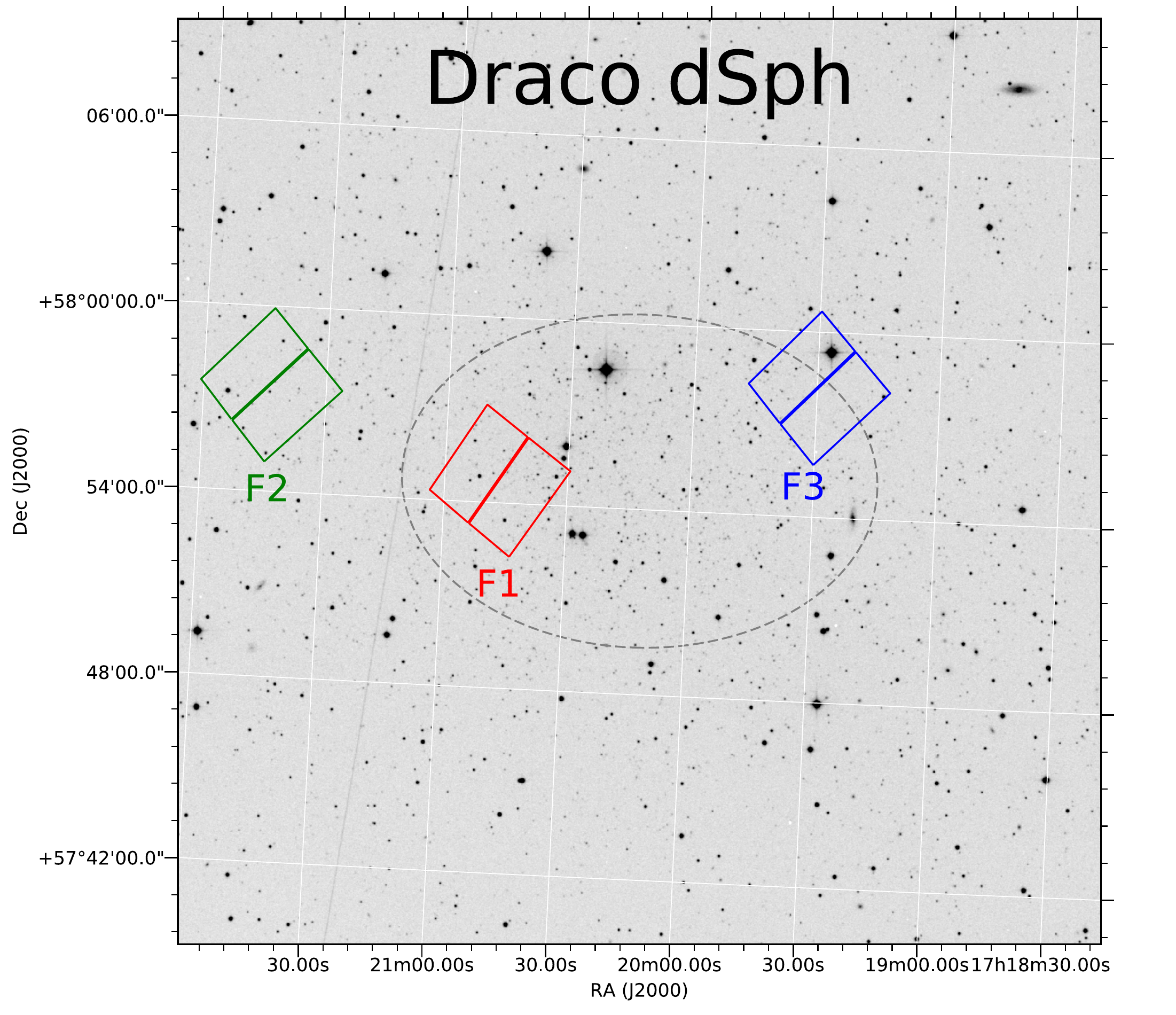}{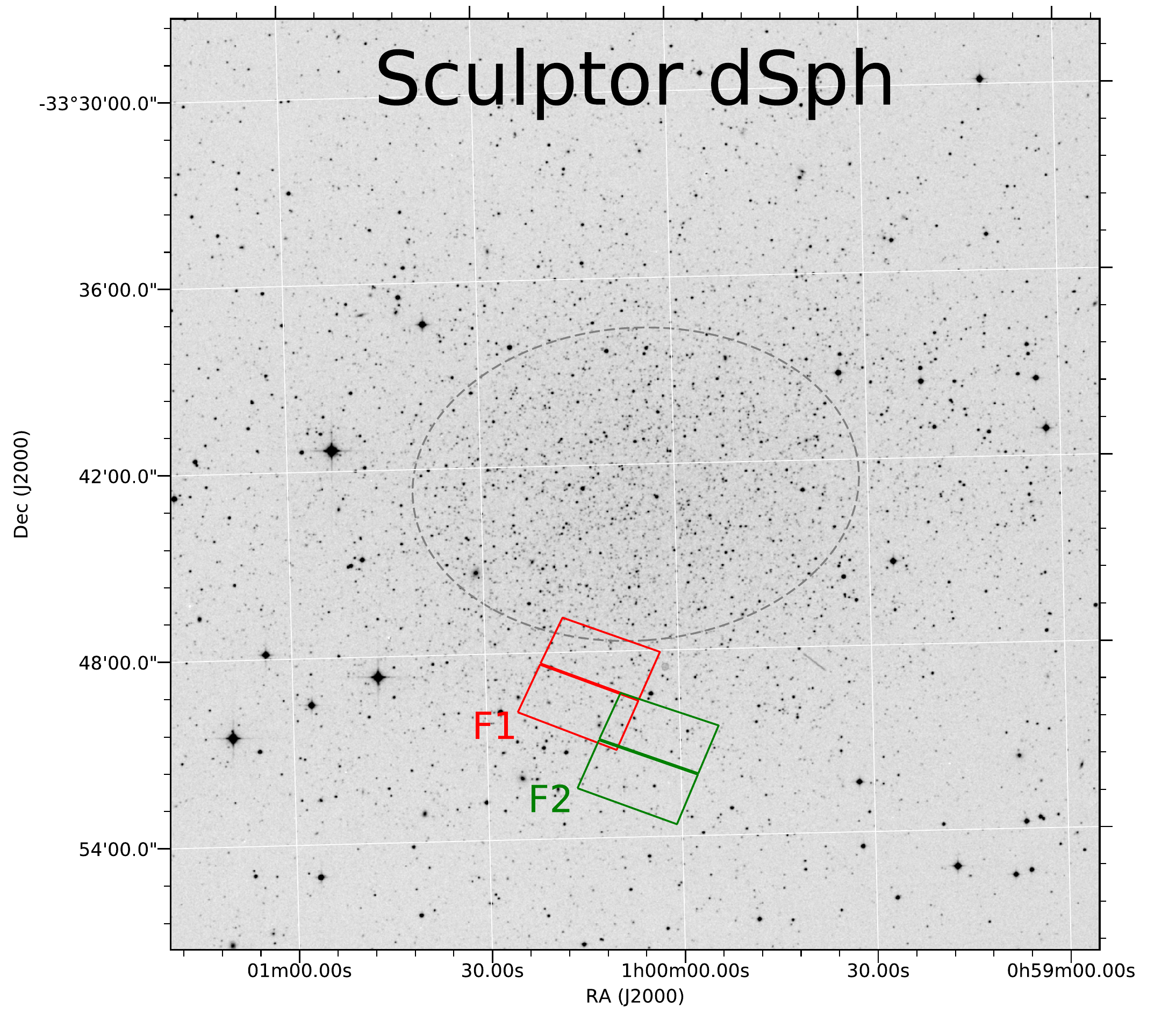}
\caption{Field locations of our ACS/WFC observations for 
         Draco (left panel) and Sculptor (right panel) plotted over a 
         30\arcmin$\times$30\arcmin\ section of the sky centered 
         on each galaxy from the STScI Digital Sky Survey. 
         The line that bisects each ACS/WFC field is a small 
         gap between the two CCDs; the CCD readout direction is roughly 
         perpendicular to this. Dashed ellipses represent core radii as 
         derived by \citet{ode01} for Draco, and by \citet{wes06} 
         for Sculptor.
         \label{f:fields}}
\end{figure*}

Figure~\ref{f:fields} shows the \hst\ ACS/WFC fields we used for  
measuring the PMs of Draco and Sculptor.
The first-epoch ACS/WFC data for Draco were observed in 2004 October 
through the \hst\ program GO-10229 (PI: S. Piatek). 
\footnote{Our {\it DRACO-F1}, {\it -F2}, and {\it -F3} fields are 
identical to the Dra~2, Dra~3, and Dra~1 fields of \citet{pry15}, 
respectively.} Field {\it DRACO-F1} was observed with ACS/WFC again 
two years later in 2006 October through \hst\ program GO-10812 to measure 
the PM of Draco. Results using these two-year baseline data have 
been reported in \citet{pry15}. The {\it DRACO-F1} and {\it -F2} fields 
were observed in F606W, while the {\it DRACO-F3} field was observed in 
F555W to avoid saturating the quasi-stellar objects (QSOs). Due to the 
failure of ACS/WFC in 2006--2007, fields {\it DRACO-F2} and {\it -F3} 
were observed with WFPC2 in 2007 December. However, we did not consider 
using the WFPC2 data for PM measurements for the same reasons as 
discussed in our Leo~I paper \citep{soh13}. 

All three Draco fields were re-observed through our \hst\ program 
GO-12966 (PI: R. van der Marel) using the same configurations 
(i.e., filters, telescope pointings, and orientations) as in the 
2004--2006 observations. The {\it DRACO-F1} field was observed in 
2013 October, and {\it DRACO-F2} and {\it -F3} in 2012 October.
All three target fields of Draco have QSOs in them as well as 
plenty of bright and compact background galaxies that can be 
used as stationary reference objects.

For Sculptor, we used two fields just outside the core radius as 
shown in Figure~\ref{f:fields}. The first-epoch data for Sculptor
were observed in 2002 September through \hst\ program GO-9480 to 
measure the weak lensing (or cosmic shear) of background galaxies.
We re-observed these two fields in 2013 September, again using the 
same telescope pointings and orientations as in the 2002 observations.

In the course of our second-epoch observations through program 
GO-12966, we also acquired short exposures in different filters 
(F814W for the Draco fields, and F606W for the Sculptor fields) 
to construct color-magnitude diagrams (CMDs) of stars in our 
target fields. A summary of observations for each target galaxy 
is shown in Table~\ref{t:obslog}.

The primary goal of our \hst\ GO-12966 program is to study the 
internal PM dynamics of stars in Draco and Sculptor, and we are in 
the process of analyzing the results which will be presented in a 
separate forthcoming paper. All of the exposures obtained through our 
\hst\ GO-12966 program made use of the experimental POST-FLASH 
capability to mitigate the impact from charge transfer efficiency 
(CTE) losses. This was important because the typical exposure time 
for individual images were all about 500 sec, which is significantly 
less than those in our other studies \citep[e.g.,][]{soh13}.

%%%%%%%%%%%%%%%%%%%%%%%%%%%%%%%%%%%%%%%%%%%%%%%%%%%%%%%%%%%%%%%%%%%%%%
%% FIGURE 1
%%%%%%%%%%%%%%%%%%%%%%%%%%%%%%%%%%%%%%%%%%%%%%%%%%%%%%%%%%%%%%%%%%%%%%
%
% 
%%%%%%%%%%%%%%%%%%%%%%%%%%%%%%%%%%%%%%%%%%%%%%%%%%%%%%%%%%%%%%%%%%%%%%

\subsection{Measurements}\label{ss:measurements}

We compared the two epochs of F606W/F555W (for Draco) and F775W (for Sculptor) 
observations to measure the absolute PMs of our target galaxies. This 
was accomplished by determining the shifts of member stars in the dSphs 
with respect to two different types of stationary objects, galaxies and QSOs, 
in the distant background. Our methodology generally follows that of our 
previous works on M31 and Leo~I \citep{soh12,soh13}, and so we refer 
readers interested in the the details to those papers. Here we outline 
the main features of our PM derivation process. 

\subsubsection{Initial Analysis Steps}

We downloaded both the regular flat-fielded {\tt \_flt.fits} and corrected 
{\tt \_flc.fits} images from the Mikulski Archive for Space Telescopes (MAST). 
The latter images are pre-processed for the imperfect charge transfer 
efficiency (CTE) using the pixel-based correction algorithms of \citet{and10}. 
The PM measurements were performed using both sets of images since we were 
uncertain how the current version of the CTE correction routine we used 
performs on images obtained using the POST-FLASH option. We found that the 
{\tt \_flc.fits} images taken with the POST-FLASH option were somewhat 
overcorrected for the imperfect CTE which causes systematics in our PM 
measurements. Therefore, the final results were all derived from the 
{\tt \_flt.fits} images. We carefully examined the individual 
{\tt \_flt.fits} images for both epochs and found that the level of CTE 
loss for the images taken in 2012--2013 are comparable to those of 
the 2004--2006 data thanks to the 2012-2013 POST-FLASH observations. 
In the end, this has worked to our advantage for PM measurements since 
the impact of CTE loss on astrometry was found to almost cancel out
when taking the difference in positions of objects between the two epochs.
As we discuss below, we also used local corrections for mitigating the 
residual systematics.

As the first step, we processed the {\tt\ \_flt.fits} images using the 
{\tt img2xym\_WFC.09x10} program from \citet{and06} to obtain a position 
and a flux for each star in each exposure. We applied corrections to 
the the positions using the known ACS/WFC geometric distortions. 
We then created a high-resolution stacked image for each field using the 
first- (for the Draco dSph) and second-epoch (for the Sculptor dSph) 
images. Stars and galaxies were then identified from the stacked images.
Photometric measures from the {\tt img2xym\_WFC.09x10} program 
were used to create a CMD for each field, and this was used to 
identify member stars of our target dSphs. The subsequent analysis 
steps are different depending on which type of background objects are 
used as stationary references sources. We discuss further steps for 
each case below.

\subsubsection{Background Galaxies as Stationary Reference Sources}
\label{sss:galaxies}

For each star and background galaxy, a template was constructed from 
the high-resolution stacked image. This template was used to measure 
a position for each object in each individual exposure in each epoch.
Templates were fitted directly to the images of the epoch from which 
they were created (first-epoch for Draco, and second-epoch for Sculptor).
For fitting templates to the images of the other epoch, we included 
$7\times7$ pixel convolution kernels to allow for PSF differences 
between epochs. These kernels were derived using bright and isolated 
Draco/Sculptor stars distributed throughout the fields.

The template-based positions of stars for multiple exposures were 
averaged and used to define first- (for the Draco dSph) or second-epoch 
(for the Sculptor dSph) reference frames. We used the positions of 
the stars in each of the second- (Draco) or first-epoch (Sculptor) 
exposures to transform the template-measured positions of the galaxies 
into the reference frames. We then took the difference between the 
first- and second-epoch positions of galaxies to obtain the relative 
displacement of the galaxies with respect to the dSph stars. To remove 
any remaining systematic PM residuals associated with the detector 
position and brightness of sources (e.g., due to imperfect CTE 
corrections) we derived and applied a local correction for each 
background galaxy using stars of similar brightness that lie in 
the vicinity. Finally, we multiply the relative displacements of the 
galaxies by $-1$ to obtain the mean absolute displacement of the dSph 
stars, since in reality the galaxies/QSOs are stationary and the stars 
are moving. Multiplying the resulting displacements by the pixel scale 
of our reference images ($50\>{\rm mas}\,{\rm pix}^{-1}$), and dividing 
by the time baseline turns our results into actual PMs.

For the {\it DRACO-F1} field, we have data obtained in 2004, 2006, and 
2013 as described in Section~\ref{ss:data}. For the final PM measurement, 
we used the 2004 and 2013 data as our first and second epoch, respectively. 
The 2006 data were used to provide an extra check (see 
Section~\ref{sss:results_draco}), but they were not included in our 
final results.

Because the {\it DRACO-F3} field was observed with F555W, which has only 
about half the bandwidth of F606W, the background galaxies are significantly 
fainter than those detected in the F606W images. After attempting to 
construct and fit templates to the background galaxies in this field, 
we concluded that the overall quality of positional measurements were 
too poor to include in our results. For this reason, the PM results for 
{\it DRACO-F3} field are only reported using QSOs as stationary reference 
sources (see Section~\ref{sss:qso}).

\subsubsection{QSOs as Stationary Reference Sources}
\label{sss:qso}

All three of our Draco fields include QSOs in them, and we use these 
objects to provide an independent measurement of the Draco PM.  
For the {\it DRACO-F1} field, \citet{pry15} used two QSOs, one detected 
in the top ACS/WFC chip (WFC1), and the other detected in the bottom chip 
(WFC2). Both QSOs were easily identified in our images thanks to Figure~2 
of \citet{pry15}. However, due to the increase in individual exposure times 
for our GO-12966 data, we found that the QSO located in WFC1 is slightly 
saturated in the images taken in 2012, making its positional measurement 
unreliable. We therefore decided to only use the QSO detected in WFC2. 
This QSO, and the QSOs in the other two Draco fields were detected in the 
2012 data with counts well below the saturation limits.

For measuring PMs using QSOs as reference sources, we used the positions 
of QSOs and stars measured based on the library PSFs by the 
{\tt img2xym\_WFC.09x10}, instead of using the template-based positions 
described in Section~\ref{sss:galaxies}. This is because the PSFs of QSOs 
are virtually indistinguishable from the PSFs of stars, and because the 
library-based positions are more accurate than the template-based 
positions. We start by only selecting member stars of Draco, based on 
their CMD properties, that are detected on the same ACS/WFC image quadrant 
as the QSO. The positions of these stars in each individual image are 
corrected for the known geometric distortions, and subsequently averaged 
separately for the first- and second-epoch data. The positions of stars 
in the first epoch data are used to define a reference frame. We then used 
the positions of stars in the second epoch to transform the position of 
the QSO into the reference frame. As a result, the PM of Draco stars can be 
inferred by taking the difference between the first-epoch reference QSO 
position and the second-epoch transformed QSO position, multiplying the 
results by $-1$, converting pixels to mas, and dividing by the time 
baseline. 

The QSOs we used for measuring PMs are typically brighter than most of 
the Draco member stars we used for setting up the reference frame. 
For example, in the same quadrant as the QSO located in the {\it DRACO-F1} 
field, there are only 18 out of 132 Draco stars that are brighter than 
the QSO. For the other two fields, the situation is worse: only two and 
one out of 68 and 42 Draco stars are brighter than the QSOs in fields 
{\it DRACO-F2} and {\it F3}, respectively.
This can potentially cause CTE-related systematics since the CTE degradation 
is known to be a strong function of the brightness of a source, and we 
are using stars of different brightnesses than the QSOs to define the 
reference frame. To correct for this effect, the procedure described above 
was iterated using Draco stars in different brightness ranges to define 
the reference frame. In our first iteration, the measurement was 
carried out using stars brighter than an instrumental magnitude 
of $m_{\rm instr} = -9.00$. 
\footnote{The instrumental magnitude is defined as $m_{\rm instr} = 
-2.5 \log (cnts)$, where $cnts$ is the number of counts in 
the $5\times5$ pixels around the brightest pixel.}
In subsequent iterations, we decreased this limit in steps of 0.5 mags 
until the faint limit was $m_{\rm instr} = -11.50$. For each step, we 
compute the PMs of the QSOs and the median magnitude of stars used in 
the transformation process. This provides a relation between the 
brightness of stars and the measured PMs. We fit a line to this relation, 
and computed the PM for the case of stars having the same brightness 
as the QSOs. These relations are monotonic implying that we are indeed 
correcting for the residual CTE effect. The final PMs of Draco stars 
with respect to the stationary QSOs were then derived using the same 
procedure as outlined in Section~\ref{sss:galaxies}. 
The final PM uncertainties were obtained by taking the quadrature sum 
of the uncertainties in the average positions at the two epochs, and 
the uncertainties from fitting the lines to the PM versus brightness 
relation.

As with the case of using background galaxies as stationary references, 
we used the {\it DRACO-F1} field's 2006 data only as an extra check, and 
our final PM for this field was obtained using 2004 versus 2013 data.

\subsection{Results}\label{ss:results}

\subsubsection{Draco Dwarf Spheroidal Galaxy}
\label{sss:results_draco}
%%%%%%%%%%%%%%%%%%%%%%%%%%%%%%%%%%%%%%%%%%%%%%%%%%%%%%%%%%%%%%%%%%%%%%
%% TABLE 2
%%%%%%%%%%%%%%%%%%%%%%%%%%%%%%%%%%%%%%%%%%%%%%%%%%%%%%%%%%%%%%%%%%%%%%
%
\begin{deluxetable}{lcccc}
\tablecaption{Final Proper Motion Results for the Draco dSph.
              \label{t:dracopm}
}
\tablehead{
\colhead{}      & \colhead{$\muw$\tablenotemark{a}} & \colhead{$\mun$\tablenotemark{b}} & \colhead{$\sigma_{\muw}$} & \colhead{$\sigma_{\mun}$} \\
\colhead{Field} & \multicolumn{2}{c}{($\masyr$)}      & \multicolumn{2}{c}{($\masyr$)} 
}
\startdata
{\bf F1} (Galaxies) & $-0.0168$ & $-0.1958$ & $0.0290$ & $0.0294$ \\
{\bf F1} (QSO)      & $-0.0463$ & $-0.2025$ & $0.0188$ & $0.0164$ \\
{\bf F2} (Galaxies) & $-0.0526$ & $-0.1812$ & $0.0264$ & $0.0265$ \\
{\bf F2} (QSO)      & $-0.0825$ & $-0.1478$ & $0.0174$ & $0.0179$ \\
{\bf F3} (QSO)      & $-0.0512$ & $-0.1386$ & $0.0263$ & $0.0348$ \\
\hline
Weighted average    & $-0.0562$ & $-0.1765$ & $0.0099$ & $0.0100$ \\
\enddata
\tablenotetext{a}{Proper motion in direction of West. Note that $\muw = -\mu_{\alpha}*\cos\delta$.}
\tablenotetext{b}{Proper motion in direction of North. Note that $\mun = \mu_{\delta}$.}
\end{deluxetable}
%
%%%%%%%%%%%%%%%%%%%%%%%%%%%%%%%%%%%%%%%%%%%%%%%%%%%%%%%%%%%%%%%%%%%%%%

%%%%%%%%%%%%%%%%%%%%%%%%%%%%%%%%%%%%%%%%%%%%%%%%%%%%%%%%%%%%%%%%%%%%%%
%% FIGURE 2
%%%%%%%%%%%%%%%%%%%%%%%%%%%%%%%%%%%%%%%%%%%%%%%%%%%%%%%%%%%%%%%%%%%%%%
%
\begin{figure*}
\plottwo{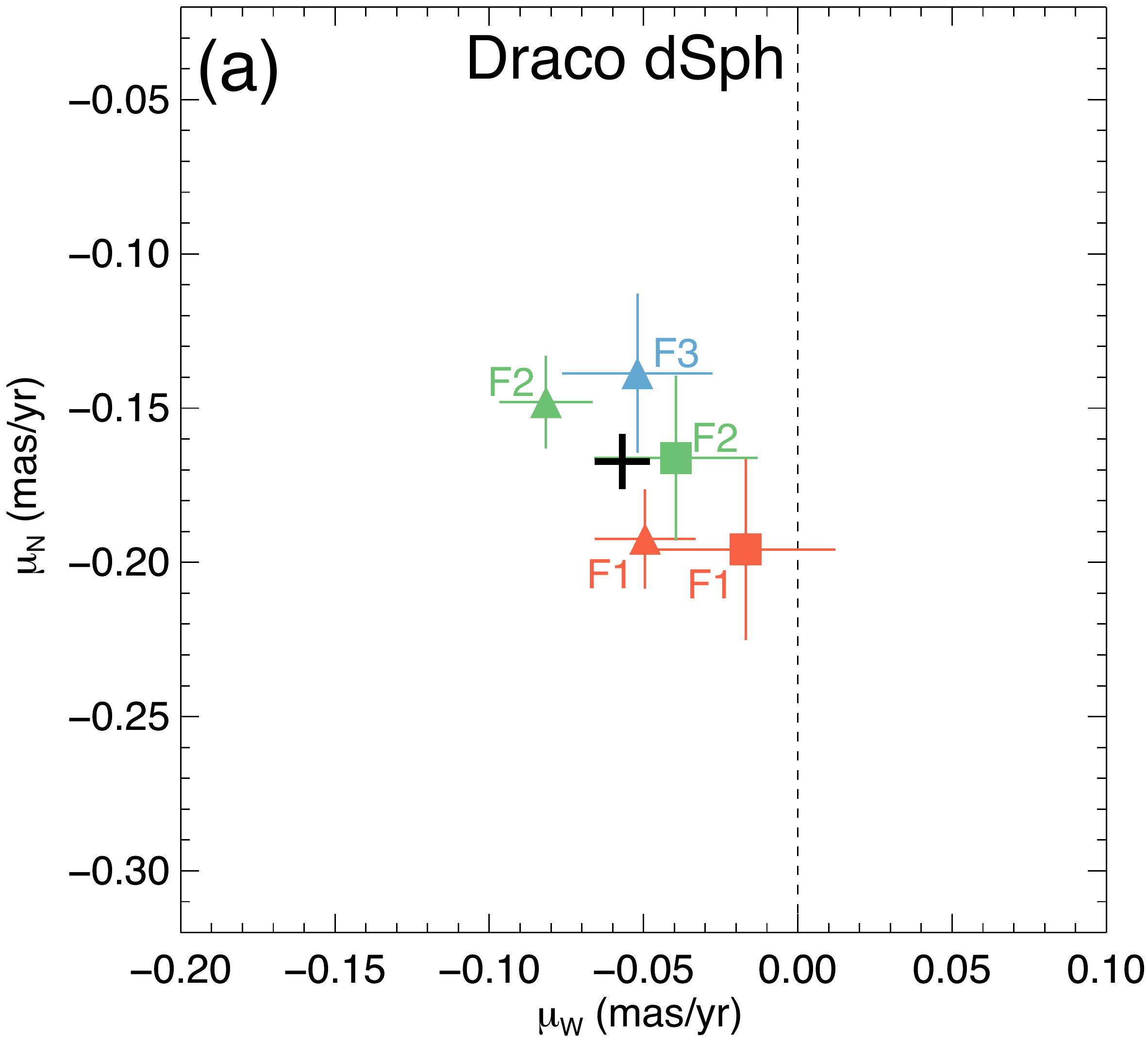}{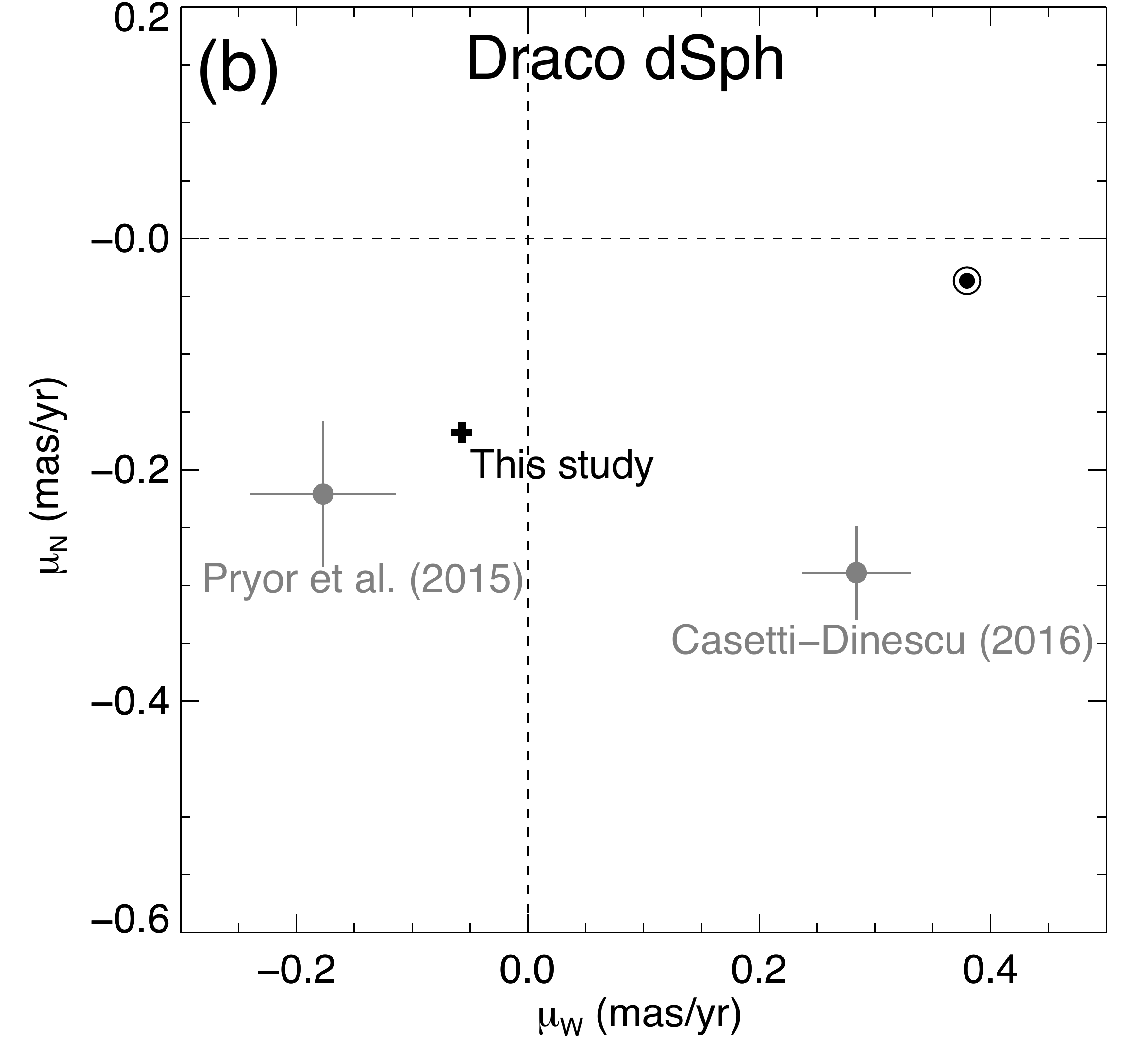}
\caption{Proper motions results in $(\muw, \mun)$ for the Draco dSph. 
         The origins correspond to the velocity such that Draco has no 
         transverse motion in the heliocentric rest frame.
         In panel (a), each square or triangle with error bars indicates 
         a PM measured using background galaxies or QSOs as stationary 
         references, respectively (see the text for details). Data 
         points with different colors are for measurements from 
         different fields as labeled in the figure. The black plus 
         symbol represents our final weighted average of the five 
         individual measurements and its uncertainty. In panel (b), we 
         compare our PM results (black plus symbol) with those from 
         two recent studies \citep{pry15,cas16} as labeled. The solar 
         symbol corresponds to the velocity such that Draco has no 
         tangential velocity in the Galactocentric rest frame. 
         \label{f:dracopm}
        }
\end{figure*}
%
%%%%%%%%%%%%%%%%%%%%%%%%%%%%%%%%%%%%%%%%%%%%%%%%%%%%%%%%%%%%%%%%%%%%%%

Our PM results for the Draco dSph are presented in Table~\ref{t:dracopm}, 
and the corresponding PM diagram is shown in Figure~\ref{f:dracopm}a.
PM measurements using different background sources are plotted in different 
symbols, and results from each field are plotted in different colors.
The PM results for the {\it DRACO-F1} field in Table~\ref{t:dracopm} was 
derived using a data set with a time baseline of 9 yr (2004 versus 2013).
However, since we have images acquired in 2006 for this field, we used 
them as an external check by measuring PMs of Draco stars using 2006 
data as the first epoch, and 2014 data as the second epoch. We followed 
the same procedure outlined in Section~\ref{ss:measurements} to obtain 
PMs using both QSO and background galaxies as stationary references.
The resulting 7 yr-baseline PMs are consistent within $1.5\sigma$ 
compared to the 9 yr-baseline PMs listed in the first two lines of 
Table~\ref{t:dracopm} with slightly larger uncertainties as expected 
from the shorter time baseline.\footnote{We obtain $(\mu_{W},\>\mu_{N}) 
= (-0.0264 \pm 0.0385,\>-0.2141 \pm 0.0396) \masyr$ using background 
galaxies, and $(-0.0709 \pm 0.0246,\>-0.1695 \pm 0.0222) \masyr$ using 
QSO as stationary references.} This provides an additional check on our 
PM results for the {\it DRACO-F1} field.

The uncertainties in the measurements are dominated by the random 
errors in the reference frame set by background galaxies or QSOs.
These random errors are independent from each other. We therefore  
calculate the average PM of Draco by taking the error-weighted 
mean of the five measurements provided in Table~\ref{t:dracopm},
which yields  
\begin{equation}
\label{e:dracopm}
  (\mu_{W},\>\mu_{N}) =
  (-0.0562 \pm 0.0099,\>-0.1765 \pm 0.0100)\ {\rm mas\ yr}^{-1} .
\end{equation}
The final average of the five data points and associated uncertainties 
in each coordinate are plotted as a black cross in Figure~\ref{f:dracopm}a. 

Overall, we find that measurements using different objects as stationary 
references agree well with each other. This provides confidence on our 
Draco PM results, and more generally on the PM measurement technique 
using background galaxies as stationary objects. Measurements for 
different fields also agree to within the error bars. 
To test the statistical agreement among the individual measurements 
listed in Table~\ref{t:dracopm}, we calculate the quantity 
\begin{equation}
\label{e:chisq}
  \chi^2 = \sum_{i} 
     \left[
     \left ( \frac{ \mu_{W,i} - {\overline \mu_{W}} }
               { \Delta \mu_{W,i} } \right )^2 
     +
     \left ( \frac{ \mu_{N,i} - {\overline \mu_{N}} }
               { \Delta \mu_{N,i} } \right )^2
     \right].
\end{equation}
This quantity is expected to follow a probability distribution 
with an expectation value of the number of degrees of freedom ($N_{\rm DF}$)
with a dispersion of $\sqrt{2N_{\rm DF}}$. Since we have five independent 
measurements each in two directions on the sky, the $\chi^2$ is then 
expected to have a value of $8 \pm 4$. From Table~\ref{t:dracopm} and 
Equation~\ref{e:chisq}, we find $\chi^2 = 11.1$. Therefore, we find
that our measurements in Table~\ref{t:dracopm} are consistent within 
our quoted uncertainties. 

Our final 1d PM uncertainty for Draco is $10\musyr$ in each direction. 
This is smaller than any other measurement uncertainties we have achieved 
using our PM measurement techniques, and therefore may appear to be beyond 
\hst's astrometric capabilities. However, this small uncertainty is mainly 
due to the time baselines being longer than our previous studies. 
For example, in our M31 study \citep{soh12}, 
we achieved a 1d PM uncertainty of $\sim 12 \musyr$ for time baselines 
of 5--7 years averaging results from three separate fields. Our Draco 
data also consists of measurements from three separate fields, but the 
time baselines are about 2.5 years longer than the M31 work. Simply 
scaling uncertainties of the M31 work by this time baseline ratio gives an 
estimated uncertainty of $8.5 \musyr$, which is consistent with our measured 
uncertainty for Draco. In our Leo~I study \citep{soh13}, we achieved a 1d PM 
uncertainty of $29\musyr$ using two epochs of ACS/WFC data separated by 5 yrs 
for a single field. Scaling by the time baseline ratio, and dividing 
by $\sqrt{3}$ to account for the difference in the number of fields used 
for the measurement gives $10\musyr$, which again is consistent with our 
PM uncertainty for Draco. We conclude that our PM measurement uncertainties 
for Draco are in line with expectations from our previous studies. 

In both our previous studies mentioned above, we carried out detailed 
analyses to argue that there were no systematic errors in excess of 
the random errors. Since our smaller random errors here are merely 
due to a longer time baseline (which reduces random and systematic 
errors equally) and the higher number of fields, the same conclusions 
should hold true. Nevertheless, the difference between QSO and 
background galaxy results for {\it DRACO-F1} and {\it -F2} may indicate 
a small systematic effect. This is likely a problem with the QSO 
measurements since they (1) sample only one region on the detector, 
(2) do not average over multiple background sources, and (3) 
require a magnitude correction as demonstrated in Section~\ref{sss:qso}.
However, systematics are a problem only if they are correlated 
between different measurements, but we find no evidence for any 
such effects. Therefore, upon averaging, these systematics 
should decrease as $\sqrt(N)$ as we have assumed in the averaging 
of our results.

As shown in Figure~\ref{f:fields}, our field locations for measuring 
the absolute PM are offset from the center of Draco by angular distances 
of 12--30\arcmin. If the internal motions of Draco stars in tangential 
directions are significantly large, our PM measurement may not represent 
the center of mass (COM) motion of Draco. This is particularly true 
if the tangential motions show a systematic pattern (e.g., clockwise 
or counter-clockwise rotation). To check this, we have subtracted the 
average PM of Draco from the PMs of each of our target fields and 
plotted the residual 2d motions on the sky in the left panel of 
Figure~\ref{f:2dmotion}. We do not detect any rotational sign from 
the residual motions. 

We also carried out additional checks as follows. Whereas so far, 
there are no internal PM measurements of Draco, the line-of-sight (LOS)
velocity shows a slight rotation sign at the level of $6 \kms$ at a radius 
of 30\arcmin\ along Draco's major axis \citep{kle01}. However, given that 
this is smaller than the central LOS velocity dispersion of Draco 
\citep[$\sigma = 9.1 \pm 1.2 \kms$;][]{wil04}, it has been claimed 
as non-significant \citep{kle02}. At the distance of Draco, $6 \kms$ is 
equivalent to $0.017 \masyr$, which is about twice the size of our 
final random error in Table~\ref{t:dracopm}. If Draco is rotating 
at this speed on the sky, our residual motions above would have shown 
systematic rotational signs, but we do not detect such sign. 
We note that Draco appears quite elongated on the sky 
\citep[$e = 0.30$;][]{ode01} suggesting that it is seen at 
high inclination. This implies that the rotation in the plane of
the sky should be less than that along the LOS. Finally, even if 
the rotational motion was systematically affecting our PM results 
for each field, the final average should represent the systemic 
tangential motions of Draco given that our three target fields sample stars 
on both sides of the dwarf near the major axis at similar angular distances. 
For the reasons stated above, we adopt our PM results in 
Equation~\ref{e:dracopm} as our final measurement for Draco.

In Figure~\ref{f:dracopm}b, we compare our new PM results with the 
two recent PM measurements using data obtained with \hst\ 
\citep{pry15} and the Subaru Telescope \citep{cas16}. As mentioned 
in Section~\ref{ss:data}, the measurement by \citet{pry15} was obtained 
using a 2-year time baseline data for the {\it DRACO-F1} field.
While the two \hst\ results are consistent within 1$\sigma$ in 
$\mun$, they are discrepant at the $\sim 2\sigma$ level in $\muw$,  
despite using the same field (albeit with a shorter time baseline), 
the same type of objects (QSOs and background galaxies), and similar 
techniques as used in this study. The source of this discrepancy is 
unclear, but it is reasonable to assume that PM results with longer 
time baselines (in this case, our results) are less subject to systematics.
The comparison with the Subaru results show even larger discrepancies.
Given that \hst\ is less prone to systematics related to atmospheric 
effects and instrumental change, we believe our results are more reliable.

\subsubsection{Sculptor Dwarf Spheroidal Galaxy}
\label{sss:results_sculptor}

%%%%%%%%%%%%%%%%%%%%%%%%%%%%%%%%%%%%%%%%%%%%%%%%%%%%%%%%%%%%%%%%%%%%%%
%% TABLE 3
%%%%%%%%%%%%%%%%%%%%%%%%%%%%%%%%%%%%%%%%%%%%%%%%%%%%%%%%%%%%%%%%%%%%%%
%
\begin{deluxetable}{lcccc}
\tablecaption{Final Proper Motion Results for the Sculptor dSph.
              \label{t:sculptorpm}
}
\tablehead{
\colhead{}      & \colhead{$\muw$} & \colhead{$\mun$} & \colhead{$\sigma_{\muw}$} & \colhead{$\sigma_{\mun}$} \\
\colhead{Field} & \multicolumn{2}{c}{($\masyr$)}      & \multicolumn{2}{c}{($\masyr$)} 
}
\startdata
{\bf F1} (Galaxies) & $-0.0368$ & $-0.1222$ & $0.0367$ & $0.0368$ \\
{\bf F2} (Galaxies) & $-0.0262$ & $-0.1428$ & $0.0254$ & $0.0263$ \\
\hline
Weighted average    & $-0.0296$ & $-0.1358$ & $0.0209$ & $0.0214$ \\
\enddata
\end{deluxetable}
%
%%%%%%%%%%%%%%%%%%%%%%%%%%%%%%%%%%%%%%%%%%%%%%%%%%%%%%%%%%%%%%%%%%%%%%

%%%%%%%%%%%%%%%%%%%%%%%%%%%%%%%%%%%%%%%%%%%%%%%%%%%%%%%%%%%%%%%%%%%%%%
%% FIGURE 3
%%%%%%%%%%%%%%%%%%%%%%%%%%%%%%%%%%%%%%%%%%%%%%%%%%%%%%%%%%%%%%%%%%%%%%
%
\begin{figure*}
\plottwo{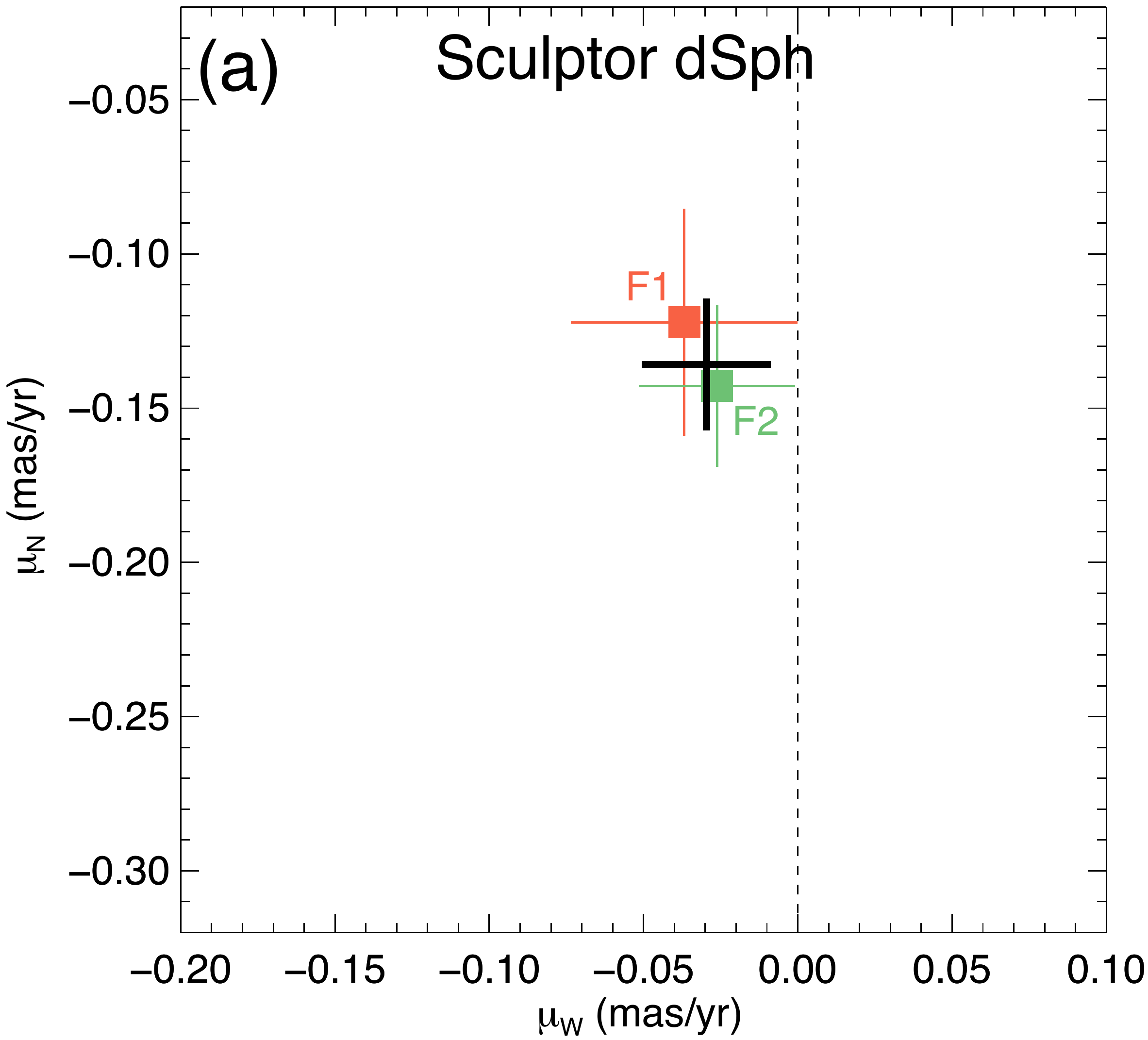}{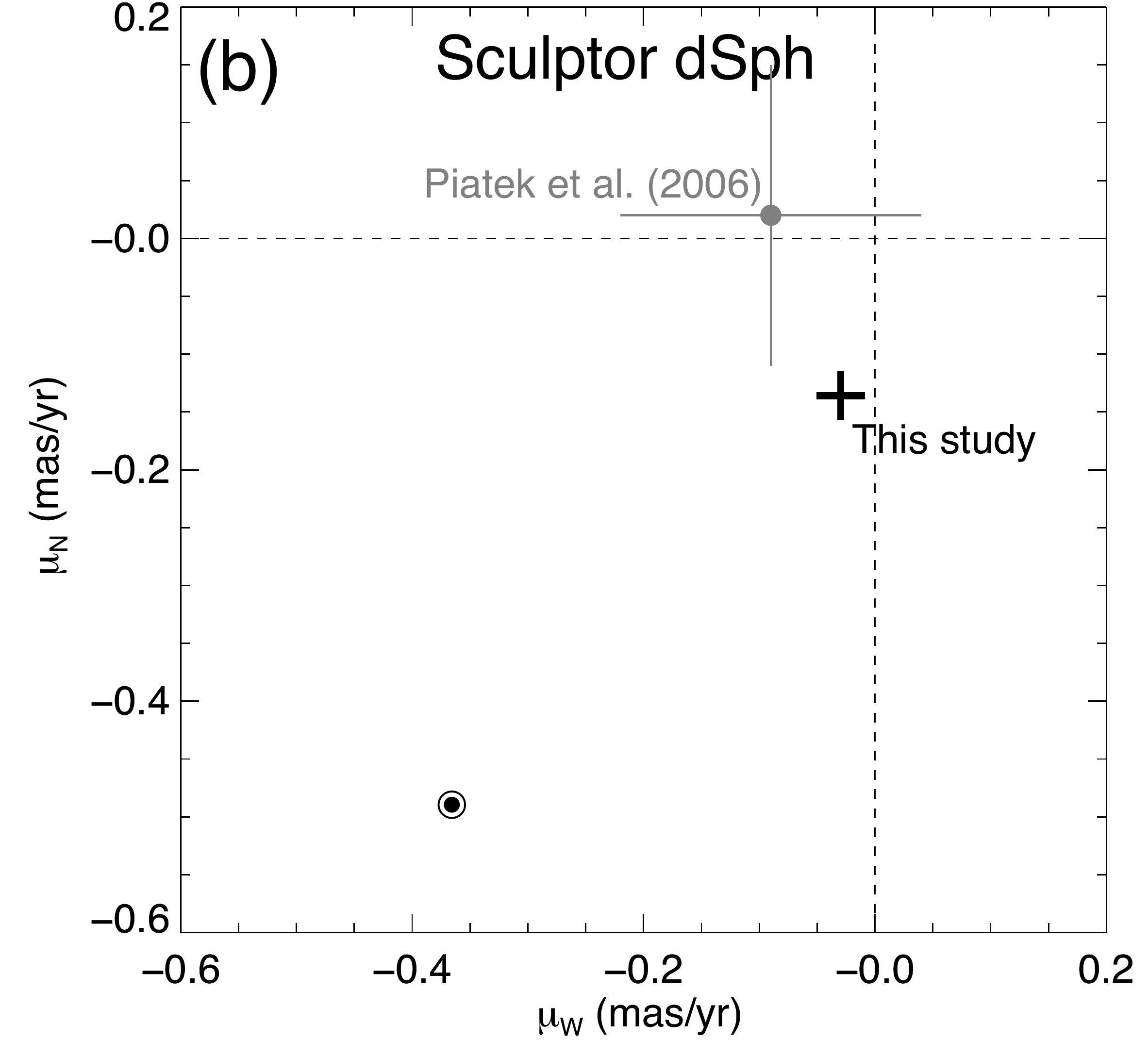}
\caption{Proper motions results in $(\muw, \mun)$ for the Sculptor dSph.
         Panels and symbols are similar to those in Figure~\ref{f:dracopm}. 
         For Sculptor, we only used background galaxies as stationary 
         references, and so each field has a single measurement 
         in panel (a). 
         \label{f:sculptorpm}
        }
\end{figure*}
%
%%%%%%%%%%%%%%%%%%%%%%%%%%%%%%%%%%%%%%%%%%%%%%%%%%%%%%%%%%%%%%%%%%%%%%

Our PM results for the Sculptor dSph are presented in Table~\ref{t:sculptorpm}, 
and the corresponding PM diagram is shown in Figure~\ref{f:sculptorpm}a.
Since we only used background galaxies as stationary references for 
this galaxy, each field has a single measurement. The error-weighted mean 
of the two fields in Table~\ref{t:sculptorpm} gives  
\begin{equation}
\label{e:sculptorpm}
  (\mu_{W},\>\mu_{N}) =
  (-0.0296 \pm 0.0209,\>-0.1358 \pm 0.0214)\ {\rm mas\ yr}^{-1}. 
\end{equation}
As evident in Figure~\ref{f:sculptorpm}a, the independent measurements 
from our two observed fields are consistent with each other within $1\sigma$.
Indeed, we find $\chi^2 = 0.3$ which is in line with the expected 
value of $2 \pm 2$. 

For Sculptor, \citet{bat08} find a radial velocity gradient 
of $7.6^{+3.0}_{-2.2}\ \kms$ per deg along its projected major axis, 
probably due to intrinsic rotation. Our target fields are located 
near the minor axis at 7--9 arcmin from the center of Sculptor. 
The residual 2d motions of our target fields after subtracting 
the average PM of Scultpor are shown as color arrows in the right 
panel of Figure~\ref{f:2dmotion}. We note that the residual motions
are too small to show compared to the average PM of Sculptor, 
demonstrating that the internal motions among the fields are 
negligible. Indeed, our 1d PM uncertainty at the distance of Sculptor 
is $8.6 \kms$, so even if we assume that Sculptor has tangential 
motions at the same level of the radial velocity gradient, our 
PM uncertainties are comparable to this. Therefore, no correction 
for the COM motion of Sculptor is required, and we adopt 
Equation~\ref{e:sculptorpm} as our final PM measurement for Sculptor.

We compare our PM results with the \hst\ measurement by \citet{pia06}. 
In their study, \citet{pia06} used QSOs in two different fields to 
measure the absolute PM of Sculptor. The two measurements agree with 
each other within $1\sigma$, with our 1d PM uncertainty being $\sim 6$ times 
smaller than that of \citet{pia06}. While both measurements employed 
the astrometric powers of \hst, \citet{pia06} used STIS data with 
time baselines of 2--3 yrs, while we used ACS/WFC data separated by 
11 yrs. Field locations are significantly different, and so 
these two measurements can be considered as completely independent. 
The agreement between the two PM measurements, despite using different 
types of background sources in different fields observed with different 
detectors, highlights the success in using \hst\ instruments as tools 
for measuring absolute PMs of dwarf galaxies in the MW halo.

\section{Space Motions}
\label{s:spacemotions}

\subsection{Systemic Motions of Draco and Sculptor on the Sky}
\label{ss:net2d}

%%%%%%%%%%%%%%%%%%%%%%%%%%%%%%%%%%%%%%%%%%%%%%%%%%%%%%%%%%%%%%%%%%%%%
%% FIGURE 4
%%%%%%%%%%%%%%%%%%%%%%%%%%%%%%%%%%%%%%%%%%%%%%%%%%%%%%%%%%%%%%%%%%%%%%
%
\begin{figure*}
\epsscale{1.15}
\plottwo{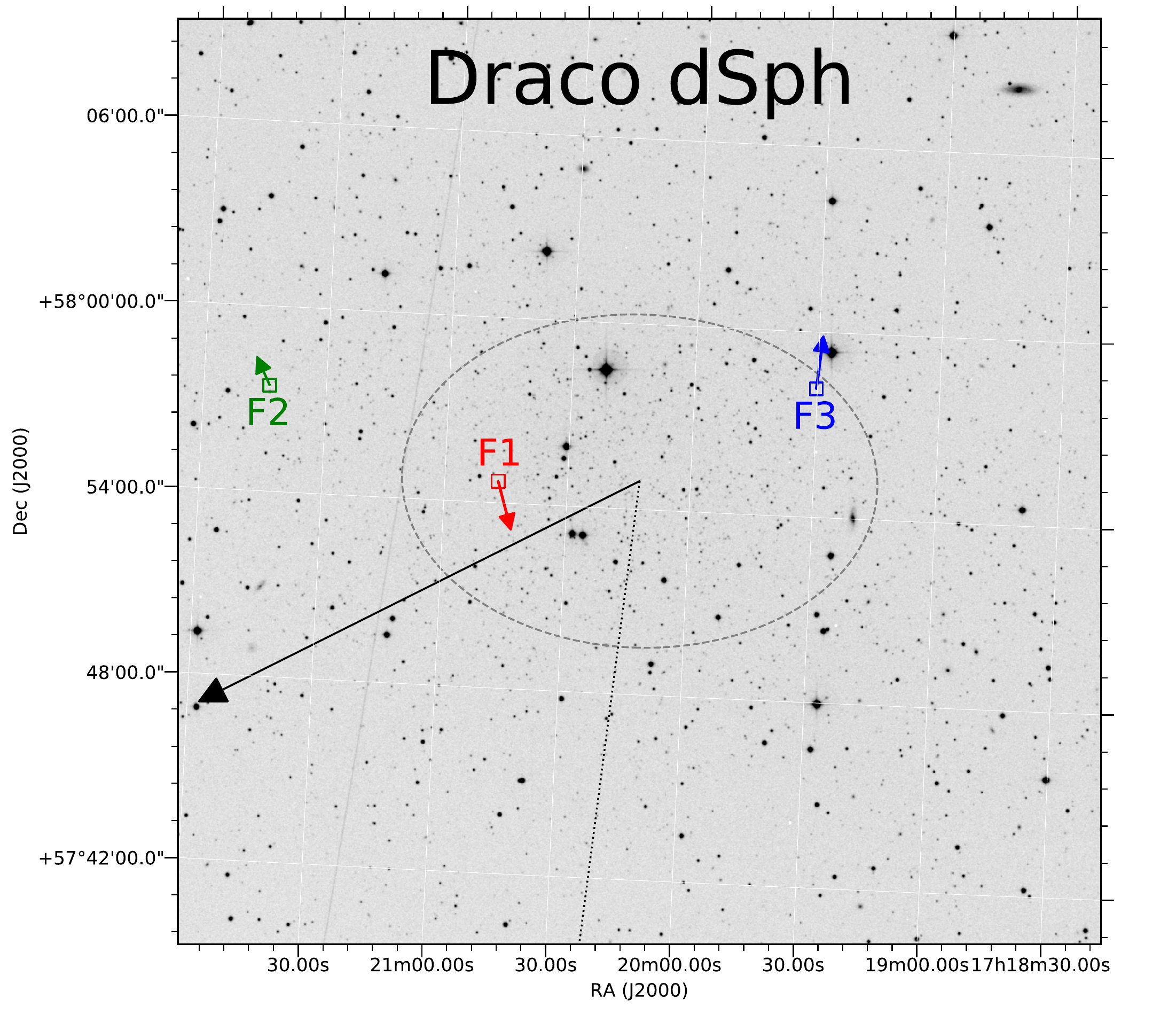}{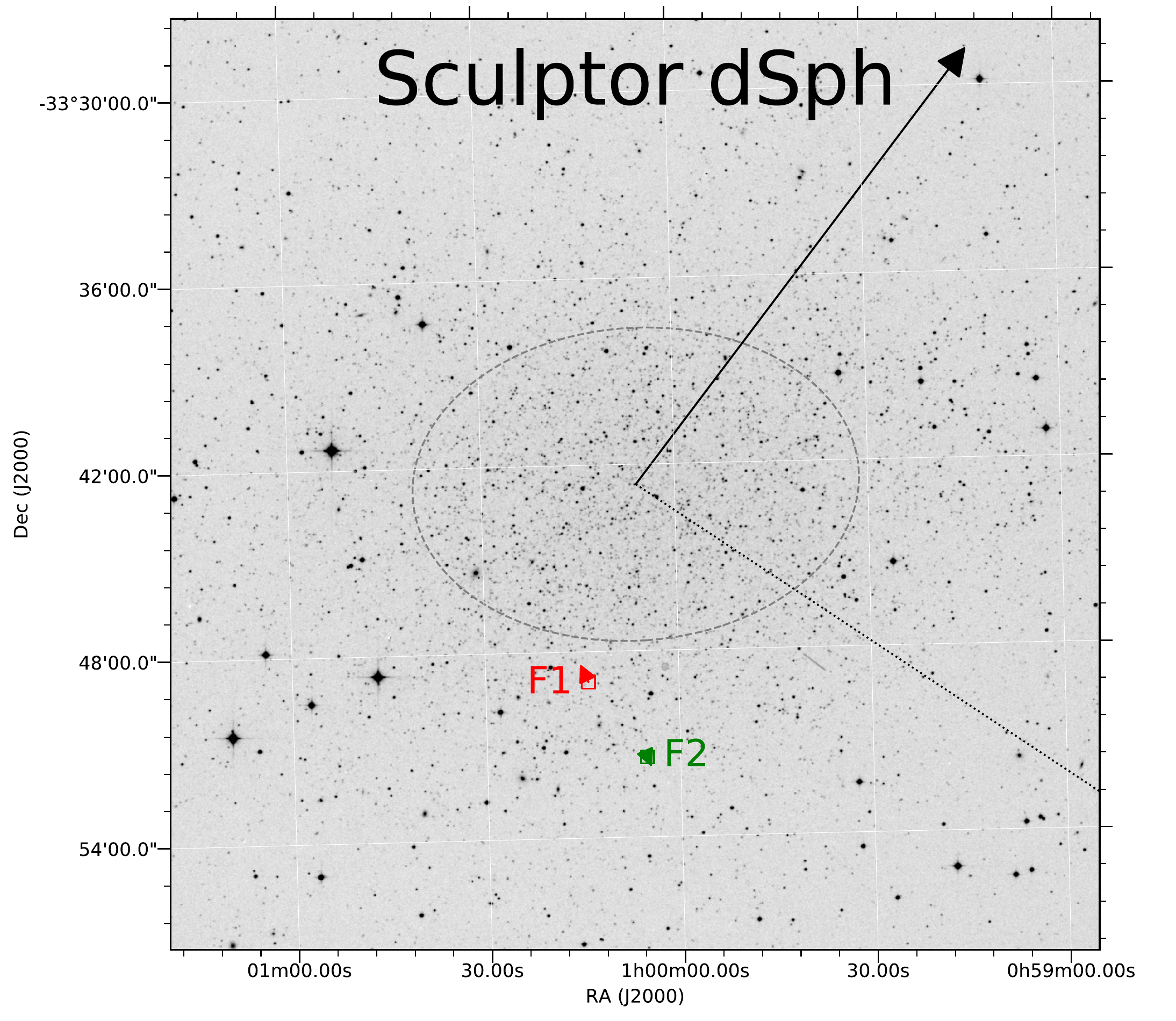}
\caption{Net average 2d motions (black arrows) and residual 2d motions 
         of target fields (color arrows) for Draco (left panel) and 
         Sculptor (right panel). For the net average 2d motions, solar 
         motions as discussed in Section~\ref{ss:net2d} have been 
         subtracted from the observed PMs. Sizes of these black 
         arrows have been arbitrarily chosen to clearly show the 
         directions of motions on the sky. For the residual motions 
         of target fields, the weighted-average PMs were subtracted 
         from the observed PM of each target field. The sizes of these 
         color arrows have been proportionally scaled with respect to 
         the black arrows to illustrate the amounts of residual 
         motions. We note that the residual motions of the Sculptor 
         fields are too small to show as arrows with reasonable length, 
         even when the black arrow was chosen to be as long as possible. 
         The directions toward the Galactic Center are indicated by 
         the black dotted lines. 
         \label{f:2dmotion}
        }
\end{figure*}
%
%%%%%%%%%%%%%%%%%%%%%%%%%%%%%%%%%%%%%%%%%%%%%%%%%%%%%%%%%%%%%%%%%%%%%%

Our PM results in Section~\ref{ss:results} include contributions from the 
motion of the Sun with respect to the MW. To obtain the systemic motions 
of Draco and Sculptor on the sky, we are required to subtract these contributions 
as follows. We adopt values of \citet{mcm11} for the Galactocentric distance 
and the rotational velocity of the Local Standard of Rest (LSR): 
$R_0 = 8.29 \pm 0.16 \kpc$ and $V_0 = 239 \pm 5 \kms$. For the solar 
peculiar velocity with respect to the LSR, we adopt values of \citet{sch10}: 
$(U_{\rm pec},\>V_{\rm pec},\>W_{\rm pec}) = (11.10,\>12.24,\>7.25) \kms$ 
with uncertainties of $(1.23,\>2.05,\>0.62) \kms$. For heliocentric distances 
to Draco and Sculptor, we adopt $76 \pm 6 \kpc$ \citep{bon04}, and $86 \pm 6 \kpc$ 
\citep{pie08}, respectively. The contributions of solar motions in 
$(\mu_{W},\>\mu_{N})$ for each dwarf galaxy is then $(0.3795,\>-0.0366) \masyr$ 
for Draco and $(-0.3657,\>-0.4895) \masyr$ for Sculptor. These are indicated 
as sun symbols in Figures~\ref{f:dracopm}b and \ref{f:sculptorpm}b.
Subtracting these solar motions from our PM measurements provides the 
net 2d motions of Draco and Sculptor on the sky: $(\muw, \mun) = 
(-0.4364,\>-0.1307) \masyr$ for Draco; and $(0.3361,\>0.3537) \masyr$ 
for Sculptor. These motions are illustrated in Figures~\ref{f:2dmotion} 
as black arrows along with the directions toward the Galactic Center as 
shown in dotted lines.

\subsection{Space Velocities in the Galactocentric Rest Frame}
\label{s:spacevel}

We adopt the same Cartesian Galactocentric coordinate system ($X,\>Y,\>Z$)
we used in our earlier studies of M31 and Leo~I \citep{soh12,soh13} to 
describe the space velocities of Draco and Sculptor.
In this system, the origin is at the Galactic Center, the $X$-axis points 
in the direction from the Sun to the Galactic Center, the $Y$-axis points 
in the direction of the Sun's Galactic rotation, and the $Z$-axis points 
toward the Galactic north pole. The position and velocity of Draco and 
Sculptor in this frame can be derived from the observed sky positions, 
distances, line-of-sight velocities, and PMs. 

\subsubsection{Draco Dwarf Spheroidal}
\label{ss:vdraco}

For Draco, the Galactocentric $(X,\>Y,\>Z)$ position is
\begin{equation}
\label{e:draco_r}
  {\vec r}_{\rm Dra} = (-4.3,\>62.3,\>43.3) \kpc.
\end{equation}
To calculate the 3-d space velocity of Draco, we adopt a heliocentric 
LOS velocity of $v_{\rm LOS} = -292.8 \pm 0.4 \kms$, estimated by 
applying the chemo-dynamical model of \citet{wal15a} to the spectroscopic 
data set of \citet{wal15b}. Combining this with our PM results in 
Section~\ref{sss:results_draco}, the Galactocentric velocity 
$(V_{\rm X},\>V_{\rm Y},\>V_{\rm Z})$ of Draco becomes
\begin{equation}
\label{e:draco_v}
  {\vec v}_{\rm Dra} = (61.0,\>16.3,\>-173.0) \pm (6.4,\>5.8,\>3.2) \kms.
\end{equation}
The uncertainties listed here and hereafter were obtained from a 
Monte Carlo (MC) scheme by propagating all observed uncertainties 
(distance, velocity, and their correlations) including those for the Sun.
The corresponding Galactocentric radial and tangential velocities are then
\begin{equation}
\label{e:draco_vradvtan}
  (V_{\rm rad},\>V_{\rm tan})_{\rm Dra} = (-88.6,\>161.4) \pm (4.4,\>5.6) \kms, 
\end{equation}
and the observed total velocity of Draco with respect to the MW is
\begin{equation}
\label{e:draco_vtot}
  V_{\rm tot, Dra} \equiv |{\vec v_{\rm Dra}}| = 184.1 \pm 4.3 \kms. 
\end{equation}

\subsubsection{Sculptor Dwarf Spheroidal}
\label{ss:vsculptor}

For Sculptor, the Galactocentric position is 
\begin{equation}
\label{e:sculptor_r}
  {\vec r}_{\rm Scl} = (-5.2, -9.8, -85.4) \kpc.
\end{equation}
We adopt a heliocentric LOS velocity of $v_{\rm LOS} = 111.5 \pm 0.3 \kms$,
obtained by applying the model of \citet{wal15a} to the spectroscopic 
data of \citet{wal09}, and combining this with our PM results for Sculptor, 
we obtain a Galactocentric velocity of 
\begin{equation}
\label{e:sculptor_v}
  {\vec v}_{\rm Scl} = (36.0,\>186.3,\>-96.7) \pm (8.8,\>10.9,\>1.3) \kms.
\end{equation}
The Galactocentric radial and tangential velocities are 
\begin{equation}
\label{e:sculptor_vradvtan}
  (V_{\rm rad},\>V_{\rm tan})_{\rm Scl} = (72.6,\>200.2) \pm (1.3,\>10.8) \kms, 
\end{equation}
and the total velocity of Sculptor with respect to the MW is
\begin{equation}
\label{e:sculptor_vtot}
  V_{\rm tot, Scl} \equiv |{\vec v_{\rm Scl}}| = 213.0 \pm 9.9 \kms. 
\end{equation}

\subsubsection{Escape Velocities}
\label{ss:vesc}
The escape velocity of a tracer object provides first-order insights  
into the enclosed mass at its distance. The escape velocity $v_{\rm esc}$ 
for a point mass $M_{\rm MW}$ is defined as
\begin{equation}
  v_{\rm esc} = \sqrt{2GM_{\rm MW}/r},
\end{equation}
where $r$ is the Galactocentric distance to the tracer object. 
According to cosmological simulations, it is unlikely to find an 
unbound satellite at the present epoch near a MW-size galaxy \citep[][but 
see Section~\ref{ss:LMC} of this paper]{boy13}. Therefore, by forcing 
Draco and Sculptor to be bound to the MW, we can use the equation above 
to calculate the lower limit on the enclosed MW mass. Using the total 
velocities from Equations~\ref{e:draco_vtot} and \ref{e:sculptor_vtot}, 
we arrive at lower limits of the enclosed MW mass $0.3\times10^{12} 
\Msun$ and $0.5\times10^{12} \Msun$ at distances of $R_{\rm GC} = 76$ 
kpc and 86 kpc, respectively.

Using the older PM measurement by \citet{pry15} and \citet{pia06}, the 
total velocities of Draco and Sculptor become $V_{\rm tot, Dra} = 225.9
\kms$ and $V_{\rm tot, Scl} = 248.1 \kms$, respectively. 
These imply lower limits of enclosed MW masses of $0.9\times10^{12} 
\Msun$ and $1.2\times10^{12} \Msun$ at $R_{\rm GC} = 76$ kpc and 86 kpc, 
respectively. In conclusion, our new PM measurements allow significantly 
lower MW masses based on the escape velocities.

\section{The Orbits of the Draco and Sculptor Dwarf Spheroidal Galaxies}
\label{s:orbits}

\subsection{Orbital Properties of Draco and Sculptor}
\label{ss:orbprops}

To explore the past orbital histories of Draco and Sculptor, we have numerically integrated 
their orbits backwards in time using the current Galactocentric positions and 
velocities derived in Section~\ref{s:spacemotions}. The orbital integration 
scheme follows the same methodology used in \citet{bes07}, \citet{soh13}, and \cite{pat17}. 

In summary, the MW's potential is modeled as a static, axisymmetric, three 
component model consisting of a dark matter halo, disk, and stellar bulge.
We adopt the same three mass models for the MW as in \citet{soh13}
with total virial masses ($M_{\rm vir}$) of 1.0$\times 10^{12}\Msun$, 
1.5$\times 10^{12}\Msun$, and 2.0$\times 10^{12}\Msun$. The MW disk mass 
was varied in each model such that the total rotation curve of the combined 
halo, disk and bulge peak at $\approx 239 \;\kms$ \citep{mcm11}. In addition, 
the MW's dark matter halo is adiabatically contracted using the {\tt CONTRA} 
code \citep{gne04}. The model parameters (concentrations, virial radii, and 
masses of the disks) for each MW model can be found in Table~2 of \citet{soh13}.

Draco and Sculptor are each modeled as Plummer spheres, with a total mass of 
$5\times10^{9}\Msun$. The softening lengths ($k_{sat}$)
are 2.3 kpc and 3.9 kpc for Draco and Sculpor, respectively. These values 
are chosen such that the halo mass matches the 
inferred total mass within the outermost data point of the empirical velocity
dispersion profile, referred to as $r_{last}$ in \citet{wal09}.

For our orbital integrations, we included the damping effects of dynamical 
friction. Since we are integrating orbits backwards in time, the damping of 
satellite orbits due to dynamical friction acts as an accelerating force. 
Dynamical friction is approximated by the Chandrasekhar formula \citep{cha43}: 
\begin{equation}
\label{eq:df}
\vec{F}_{df}= \rm - \frac{4\pi G^2 M_{sat}^2 ln \Lambda \rho(r)}{v^2} \left[ erf(X) - \frac{2X}{\sqrt\pi} exp(-X^2)\right] \frac{\vec{v}}{v},
\end{equation} 
where $X=v/\sqrt{2\sigma}$ and $\sigma$ is the one-dimensional galaxy velocity 
dispersion. Here, $\sigma$ is an approximation for an NFW profile, which was 
derived in \citet{zen03}. For three body encounters between Draco/Sculptor, 
the LMC, and the MW, the Coulomb logarithm, ln$\rm\Lambda$, takes the form 
of the 10:1 mass ratio parametrization described in \citet[][Appendix A]{vdm12b} 
for the decay of the LMC's orbit. For Draco and Sculptor, we have adopted the Coulomb
logarithm from \citet{has03}, which is $\Lambda=r/1.4k_{sat}$. For Draco and Sculptor, 
the impact of dynamical friction on their orbits is minimal. 

Following \citet{pat17}, but in contrast to \citet{bes07} and \citet{soh13}, 
the MW is not fixed in space in these calculations. Instead, the MW
moves in response to the gravitational influence of the satellites, 
particularly from the LMC (see Section~\ref{ss:LMC}), throughout the 
integration period \citep[see also][]{gom15}. 

The equations of motion corresponding to the gravitational potentials described 
above are then integrated backwards in time for 6 Gyr using a symplectic leap frog 
algorithm \citep{spr01}. Over longer timescales, the orbits of satellites are highly 
uncertain, e.g. owing to the accretion history of the MW itself \citep{lux10}.

%%%%%%%%%%%%%%%%%%%%%%%%%%%%%%%%%%%%%%%%%%%%%%%%%%%%%%%%%%%%%%%%%%%%%%
%% FIGURE 5
%%%%%%%%%%%%%%%%%%%%%%%%%%%%%%%%%%%%%%%%%%%%%%%%%%%%%%%%%%%%%%%%%%%%%%
%
\begin{figure*}
\epsscale{1.10}
\plotone{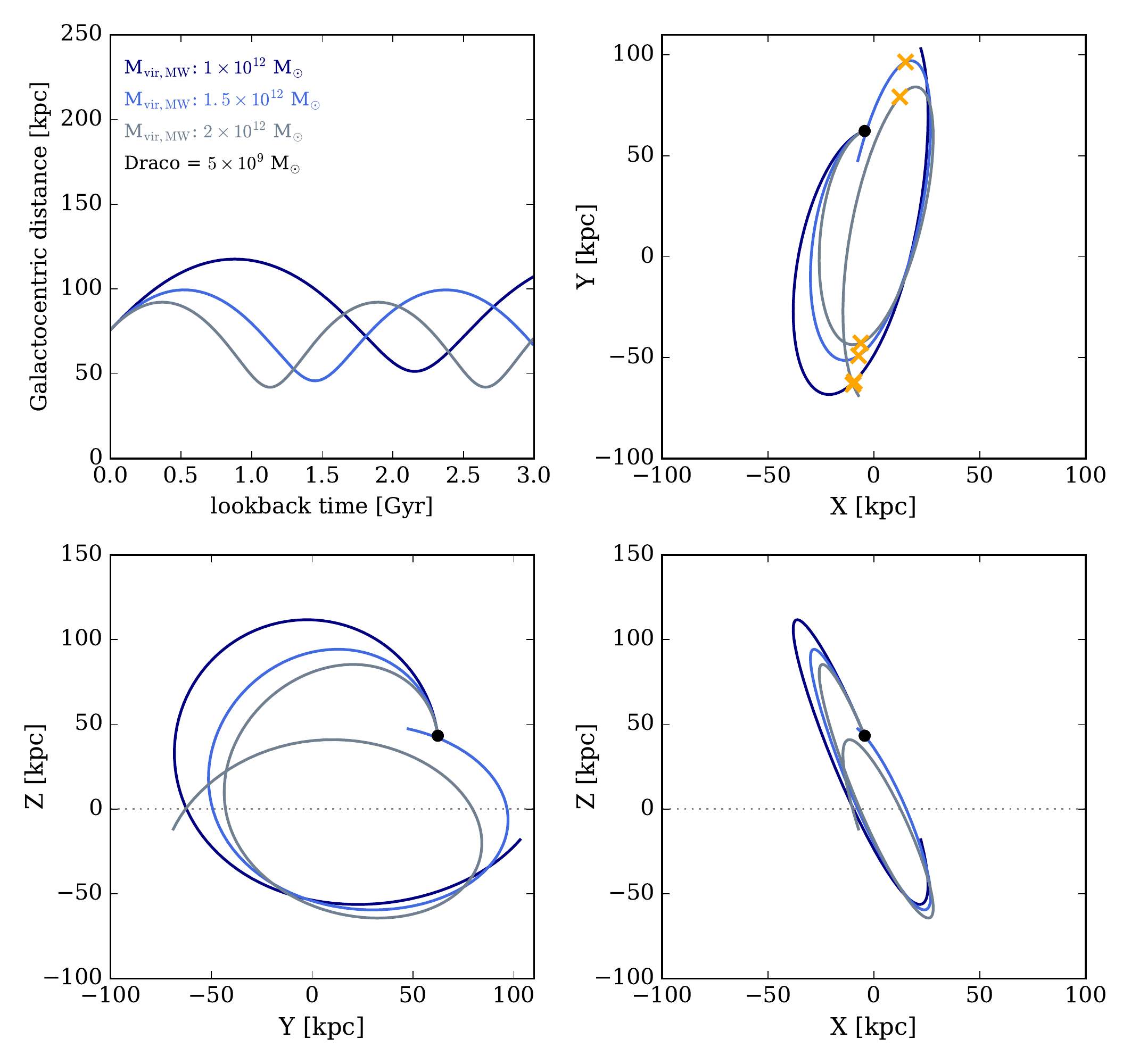}
\caption{Mean orbital history of Draco in the past 3 Gyr for the three 
different mass MW models ($M_{\rm vir} = 1.0\times10^{12} \Msun$, 
$1.5\times10^{12} \Msun$, and $2.0\times10^{12} \Msun$). The top left 
panel shows the separation between Draco and the MW as a function of 
time. The LMC is not included in these calculations. In the top right, 
bottom left, and bottom right panels, the orbital plane is presented 
in Galactocentric $X-Y$, $Y-Z$, and $Z-X$ planes, respectively. The 
current locations of Draco are indicated as black dots. The orange `x' 
markers indicate where the orbit of Draco crosses the MW's disk plane. 
The dotted gray lines in the bottom panels indicate the location of 
the MW's disk plane. 
\label{f:draco_orbit_mw}
}
\end{figure*}
%
%%%%%%%%%%%%%%%%%%%%%%%%%%%%%%%%%%%%%%%%%%%%%%%%%%%%%%%%%%%%%%%%%%%%%%

%%%%%%%%%%%%%%%%%%%%%%%%%%%%%%%%%%%%%%%%%%%%%%%%%%%%%%%%%%%%%%%%%%%%%%
%% FIGURE 6
%%%%%%%%%%%%%%%%%%%%%%%%%%%%%%%%%%%%%%%%%%%%%%%%%%%%%%%%%%%%%%%%%%%%%%
%
\begin{figure*}
\epsscale{1.10}
\plotone{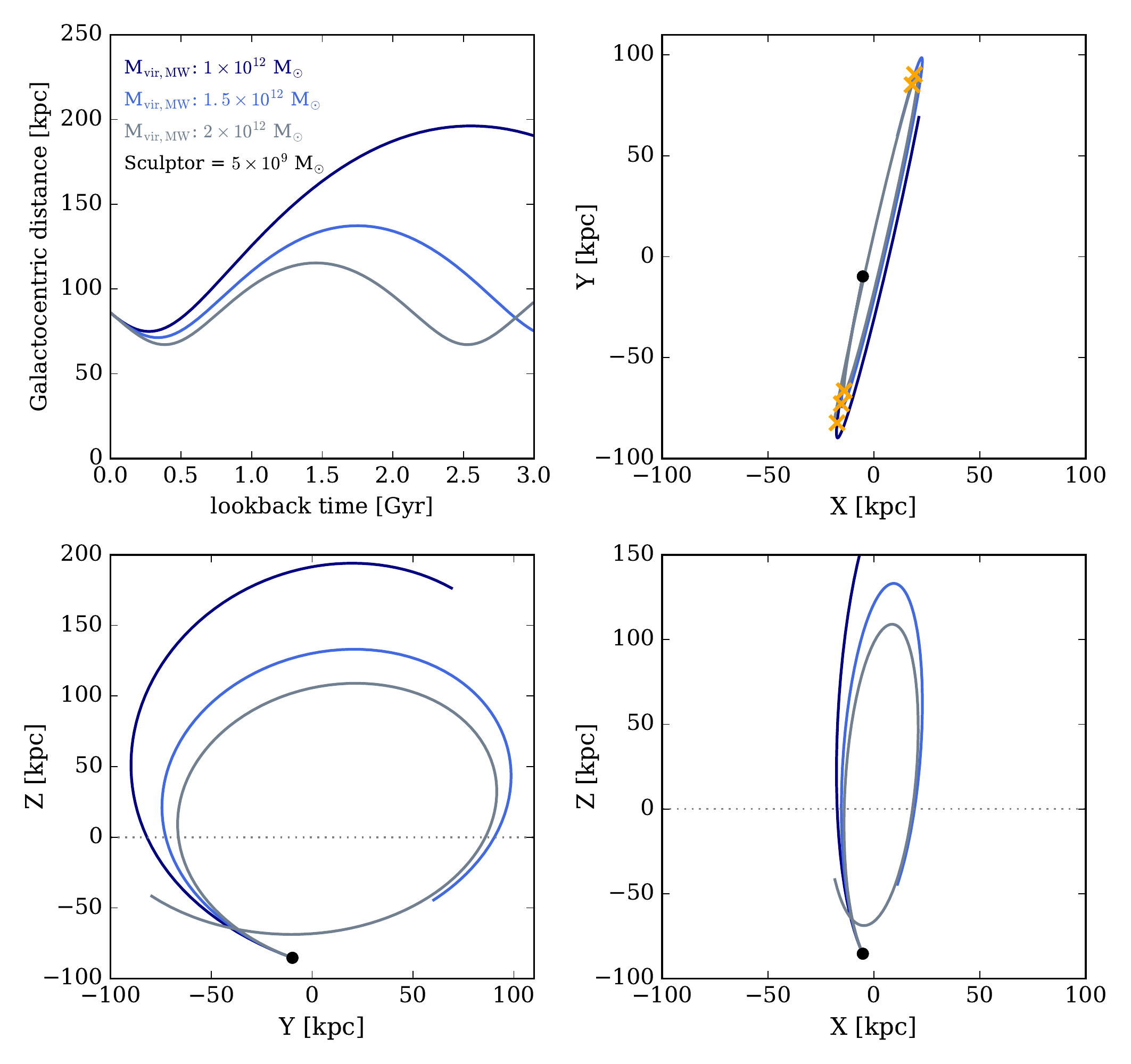}
\caption{Similar to Figure~\ref{f:draco_orbit_mw}, but now for Sculptor.
\label{f:sculptor_orbit_mw}
}
\end{figure*}
%
%%%%%%%%%%%%%%%%%%%%%%%%%%%%%%%%%%%%%%%%%%%%%%%%%%%%%%%%%%%%%%%%%%%%%%

%%%%%%%%%%%%%%%%%%%%%%%%%%%%%%%%%%%%%%%%%%%%%%%%%%%%%%%%%%%%%%%%%%%%%%
%% TABLE 4
%%%%%%%%%%%%%%%%%%%%%%%%%%%%%%%%%%%%%%%%%%%%%%%%%%%%%%%%%%%%%%%%%%%%%%
%
\begin{deluxetable*}{cccccc}
\tablecaption{Mean orbital properties of Draco and Sculptor. The 
quantities listed are the average and standard error for the most 
recent pericentric and apocentric passage. The final column lists the 
orbital period computed between the two most recent pericenters. 
All orbits for both Draco and Sculptor recover at least two 
pericentric passages and one apocentric passage, therefore the values 
for each model reflect the full set of 10,000 Monte Carlo draws from 
the $4\sigma$ error space.
\label{t:mworbits}
}
\tablehead{
\colhead{$M_{\rm MW}$}           & \colhead{$r_{\rm peri}$} & \colhead{$t_{\rm peri}$} & \colhead{$r_{\rm apo}$} & \colhead{$t_{\rm apo}$}   & \colhead{Period}  \\
\colhead{($\times10^{12}\Msun$)} & \colhead{(kpc)}          & \colhead{(Gyr)}          & \colhead{(kpc)}         & \colhead{(Gyr)} & \colhead{(Gyr)} \\
}
\startdata
\sidehead{Draco}
1.0 & $51.3\pm6.2$ & $2.2\pm0.4$ &    $121.0\pm   16.1$ & $0.9\pm0.2$ & $2.6\pm0.4$ \\
1.5 & $45.9\pm5.8$ & $1.5\pm0.2$ &    $101.3\pm   10.7$ & $0.6\pm0.1$ & $1.9\pm0.2$ \\
2.0 & $42.2\pm5.2$ & $1.2\pm0.1$ & $\phn93.4\pm\phn8.7$ & $0.4\pm0.1$ & $1.6\pm0.2$ \\
\hline
\sidehead{Sculptor}
1.0 & $74.7\pm5.2$ & $0.3\pm0.1$ &    $184.2\pm   50.5$ & $2.2\pm1.0$ & $4.7\pm0.8$ \\
1.5 & $71.0\pm5.3$ & $0.3\pm0.1$ &    $127.7\pm   25.3$ & $1.4\pm0.7$ & $2.9\pm0.3$ \\
2.0 & $66.9\pm5.3$ & $0.4\pm0.1$ &    $106.6\pm   16.4$ & $1.0\pm0.7$ & $2.2\pm0.2$ \\
\enddata
\end{deluxetable*}
%
%%%%%%%%%%%%%%%%%%%%%%%%%%%%%%%%%%%%%%%%%%%%%%%%%%%%%%%%%%%%%%%%%%%%%%

The orbital trajectories for Draco and Sculptor calculated using their 
mean positions (Equations~\ref{e:draco_r} and \ref{e:sculptor_r}) and 
velocities (Equations~\ref{e:draco_v} and \ref{e:sculptor_v}) for the 
past 3 Gyr are shown in Figures~\ref{f:draco_orbit_mw} and 
\ref{f:sculptor_orbit_mw}. The LMC is not yet included in these 
calculations. To explore the full range of plausible orbital histories, 
we use the 10,000 Monte Carlo realizations (see Section \ref{s:spacevel}), 
which sample the uncertainties in distances, radial velocities, and PMs 
from normal distributions with means and standard deviations taken from 
the observed uncertainties. We then use positions and velocities for 
each realization to integrate orbits in the three MW mass models. 
This resulted in 60,000 orbital integrations in total for Draco and 
Sculptor combined. 

Table~\ref{t:mworbits} lists the distance and look-back time 
of the most recent pericentric and apocentric passages of Draco and 
Sculptor along with the orbital period, in the case where two pericentric 
passages exist within 6 Gyr. In the majority of cases, both Draco and 
Sculptor complete multiple orbits around the MW and remain within 
its virial radius over the past 6 Gyr. Only $\sim 1\%$ of Sculptor's 
orbits in the $M_{\rm MW} = 1.0\times10^{12} \Msun$ model were in a 
``first-infall'' orbit implying that it has not completed an orbit 
about the MW. No such cases for either dwarf occur in the higher 
mass MW models, i.e., 100\% of orbits exhibit both a pericenter and 
an apocenter.

From our orbital analysis, we conclude that Draco passed the 
apogalacticon of its orbit 0.4--0.9 Gyr ago at a distance of 
$R_{\rm GC} = $93--119 kpc, and is now approaching perigalacticon 
with an orbital period of 1--2 Gyr. Sculptor, on the other hand, 
recently passed perigalacticon 0.3--0.4 Gyr ago at a distance of 
$R_{\rm GC} = $67--76 kpc, and is now moving further away from the 
Galactic center. Sculptor also has a longer orbital period of 
$\sim$2--5 Gyr. However, their average orbital eccentricities are 
similar -- both are mildly elliptical at $e \simeq 0.4$ and 
$\simeq 0.3$ for Draco and Sculptor, respectively. 
%%
%% Draco eccentricities for three different models are 0.395, 0.370, 0.373.
%% Sculptor eccentricities are 0.423, 0.285, 0.229 for three different models.
%%

In addition to being in different phases of their orbit, the two 
satellites have orbital angular momenta in almost the opposite 
direction on the celestial sphere, indicating that they orbit 
around the MW in opposite directions. This is most clearly seen 
when comparing the orbits of the two galaxies in the Y-Z plane 
(bottom left panels of Figures~\ref{f:draco_orbit_mw} and 
\ref{f:sculptor_orbit_mw}). We discuss these orbital features 
in the context of the DoS in Section~\ref{ss:dos}.

\subsection{The Dynamical Influence of the Large Magellanic Cloud}
\label{ss:LMC}

Other massive members of the Local Group may exert dynamical influence 
on the orbital histories of Draco and Sculptor. Given the distances of 
these satellites and their most likely association with the MW over 
the past $\sim$5 Gyr (see Section~\ref{ss:orbprops}), the most relevant 
perturber to their current orbital motion is the LMC. 
To examine its dynamical influence on the orbits of Draco and Sculptor, 
we added the LMC to the orbital calculations. We adopted the same 
strategy as outlined in Section~\ref{ss:orbprops} for integrating orbits 
and analyzed the three-body interactions separately for the Draco-MW-LMC 
and the Sculptor-MW-LMC systems. These orbital calculations sample the 
full 4$\sigma$ error space of the LMC's space motion and distance 
\citep{kal13}, in addition to the error space associated with Draco or 
Sculptor. Thus, each orbital realization randomly draws a set of 
position and velocity vectors from the 10,000 Monte Carlo drawings for 
the LMC and simultaneously for Draco or Sculptor. We note that the orbital 
angular momentum vector of the LMC is roughly aligned with that of Draco. 
 
\cite{gom15} showed that the orbital barycenter of the MW-LMC system 
significantly changes over time, depending on the mass of the LMC. Therefore, 
as noted earlier, the MW is not held fixed in space, but rather moves 
in response to the force of the LMC as a function of time. Our 
numerical orbit integration scheme therefore includes not only the LMC's 
gravitational torque acting on Draco and Sculptor, but also the response 
of the MW's COM to the presence the LMC. 

The LMC is modeled as a Plummer sphere, and we consider three LMC mass 
models: 0.3$\times 10^{11}\Msun$, 1.0$\times 10^{11}\Msun$, and 
2.5$\times 10^{11}\Msun$, respectively with softening lengths of 5.9, 13.1, 
and 19.5 kpc. This mass range encompasses observational constraints and 
cosmological expectations \citep[see, ][]{pat17}. 

%%%%%%%%%%%%%%%%%%%%%%%%%%%%%%%%%%%%%%%%%%%%%%%%%%%%%%%%%%%%%%%%%%%%%%
%% TABLE 5
%%%%%%%%%%%%%%%%%%%%%%%%%%%%%%%%%%%%%%%%%%%%%%%%%%%%%%%%%%%%%%%%%%%%%%
%
\begin{deluxetable*}{ccrccrcccccccc}
\tablecaption{Mean orbital parameters for the MW-LMC-Draco orbits. The 
   average distance at pericenter and apocenter, as well as their 
   corresponding times are computed using only the fraction of orbits where 
   at least one pericenter or apocenter occurs. These fractions are denoted 
   as $f_{peri}$ and $f_{apo}$. Average orbital periods are computed using 
   only the fraction of orbits where two pericenters have occurred. 
   This fraction is denoted $f_p$.
   \label{t:mwlmc_draco}
}
\tablehead{
\colhead{$M_{\rm MW}$}           & \colhead{$M_{\rm LMC}$}          & \colhead{$f_{\rm peri}$}  & \colhead{$r_{\rm peri}$} & \colhead{$t_{\rm peri}$} & \colhead{$f_{\rm apo}$} & \colhead{$r_{\rm apo}$} & \colhead{$t_{\rm apo}$} & \colhead{$f_p$} & \colhead{Period} \\
\colhead{($\times10^{12}\Msun$)} & \colhead{($\times10^{11}\Msun$)} & \colhead{(\%)}            & \colhead{(kpc)}          & \colhead{(Gyr)}          & \colhead{(\%)}          & \colhead{(kpc)} & \colhead{(Gyr)}  & \colhead{\%} &\colhead{(Gyr)}
}
\startdata
    & 0.3 & 100 & $\phn62.4 \pm \phn7.5$ & $2.7 \pm 0.5$ & 100 & $   135.5 \pm    20.4$ & $1.2 \pm 0.3$ &  99 & $3.4\pm0.6$ \\ 
1.0 & 1.0 &  96 & $\phn79.7 \pm    14.0$ & $3.8 \pm 0.8$ &  96 & $   174.7 \pm    31.8$ & $1.8 \pm 0.4$ &  77 & $4.2\pm0.7$ \\ 
    & 2.5 &   8 & $\phn98.0 \pm    24.3$ & $5.1 \pm 0.7$ &   8 & $   218.5 \pm    32.0$ & $2.4 \pm 0.4$ &   2 & $4.8\pm0.5$\\
\hline
    & 0.3 & 100 & $\phn55.4 \pm \phn7.2$ & $1.7 \pm 0.3$ & 100 & $   107.7 \pm    12.4$ & $0.7 \pm 0.1$ & 100 & $2.3\pm0.3$\\ 
1.5 & 1.0 & 100 & $\phn72.5 \pm    12.8$ & $2.2 \pm 0.4$ & 100 & $   125.5 \pm    18.4$ & $0.9 \pm 0.2$ &  99 & $3.0\pm0.8$\\ 
    & 2.5 &  93 & $   128.8 \pm    42.7$ & $4.1 \pm 0.9$ &  93 & $   201.8 \pm    43.8$ & $2.1 \pm 0.5$ &  45 & $3.8\pm0.9$ \\
\hline
    & 0.3 & 100 & $\phn50.5 \pm \phn6.4$ & $1.3 \pm 0.2$ & 100 & $\phn97.0 \pm \phn9.6$ & $0.5 \pm 0.1$ & 100 & $1.8\pm0.2$ \\
2.0 & 1.0 & 100 & $\phn65.1 \pm    10.5$ & $1.5 \pm 0.2$ & 100 & $   106.5 \pm    12.3$ & $0.6 \pm 0.1$ & 100 & $2.2\pm0.5$\\ 
    & 2.5 &  100 & $  111.7 \pm    27.9$ & $2.5 \pm 0.5$ & 100 & $   144.4 \pm    25.4$ & $1.2 \pm 0.3$ &  86 & $4.1\pm1.8$\\
\enddata
\end{deluxetable*}
%
%%%%%%%%%%%%%%%%%%%%%%%%%%%%%%%%%%%%%%%%%%%%%%%%%%%%%%%%%%%%%%%%%%%%%%

%%%%%%%%%%%%%%%%%%%%%%%%%%%%%%%%%%%%%%%%%%%%%%%%%%%%%%%%%%%%%%%%%%%%%%
%% TABLE 6
%%%%%%%%%%%%%%%%%%%%%%%%%%%%%%%%%%%%%%%%%%%%%%%%%%%%%%%%%%%%%%%%%%%%%%
%
\begin{deluxetable*}{ccrccrcccccccc}
\tablecaption{Mean orbital properties for the MW-LMC-Sculptor orbits. 
   See Table \ref{t:mwlmc_draco} for details.
   \label{t:mwlmc_sculptor}
}
\tablehead{
\colhead{$M_{\rm MW}$}           & \colhead{$M_{\rm LMC}$}          & \colhead{$f_{\rm peri}$}  & \colhead{$r_{\rm peri}$} & \colhead{$t_{\rm peri}$} & \colhead{$f_{\rm apo}$} & \colhead{$r_{\rm apo}$} & \colhead{$t_{\rm apo}$} & \colhead{$f_p$} & \colhead{Period} \\
\colhead{($\times10^{12}\Msun$)} & \colhead{($\times10^{11}\Msun$)} & \colhead{(\%)}            & \colhead{(kpc)}          & \colhead{(Gyr)}          & \colhead{(\%)}          & \colhead{(kpc)} & \colhead{(Gyr)} & \colhead{(\%)}  & \colhead{(Gyr)}
}
\startdata
    & 0.3 & 100 & $70.8 \pm 5.3$ & $0.32 \pm 0.05$ &  99 & $232.8 \pm 55.4$ & $2.9 \pm 0.9$ &  99 & $5.5\pm1.0$ \\
1.0 & 1.0 & 100 & $61.6 \pm 4.6$ & $0.35 \pm 0.04$ &  89 & $328.1 \pm 72.0$ & $4.0 \pm 1.1$ &  74 & $7.3\pm1.2$\\ 
    & 2.5 & 100 & $44.9 \pm 2.5$ & $0.32 \pm 0.02$ &  15 & $525.5 \pm 80.1$ & $5.0 \pm 0.8$ &   3 & $8.3\pm1.1$\\
\hline
    & 0.3 & 100 & $66.7 \pm 5.5$ & $0.36 \pm 0.05$ & 100 & $156.3 \pm 30.7$ & $1.7 \pm 0.7$ & 100 & $3.2\pm0.4$\\ 
1.5 & 1.0 & 100 & $58.6 \pm 5.3$ & $0.36 \pm 0.03$ & 100 & $209.3 \pm 34.7$ & $2.2 \pm 0.5$ & 100 & $3.9\pm0.5$\\ 
    & 2.5 & 100 & $47.7 \pm 4.6$ & $0.33 \pm 0.02$ &  99 & $316.3 \pm 65.2$ & $2.7 \pm 0.7$ &  99 & $5.1\pm1.4$\\ 
\hline
    & 0.3 & 100 & $62.8 \pm 5.4$ & $0.38 \pm 0.04$ & 100 & $132.8 \pm 16.3$ & $1.5 \pm 0.4$ & 100 & $2.4\pm0.2$\\
2.0 & 1.0 & 100 & $55.3 \pm 5.1$ & $0.36 \pm 0.03$ & 100 & $161.6 \pm 29.2$ & $1.5 \pm 0.5$ & 100 & $2.7\pm0.3$\\ 
    & 2.5 & 100 & $45.6 \pm 4.4$ & $0.33 \pm 0.02$ & 100 & $227.3 \pm 43.1$ & $1.7 \pm 0.5$ & 100 & $3.0\pm0.7$\\ 
\enddata
\end{deluxetable*}
%
%%%%%%%%%%%%%%%%%%%%%%%%%%%%%%%%%%%%%%%%%%%%%%%%%%%%%%%%%%%%%%%%%%%%%%

Tables~\ref{t:mwlmc_draco} and \ref{t:mwlmc_sculptor} list the distance 
and look-back time of the most recent pericentric and apocentric 
passages of Draco and Sculptor about the MW, now accounting for the 
3-body interactions of Draco/Sculptor-LMC-MW. In these tables, we also 
added columns that indicate the fraction of orbits that have a perigalactic
approach ($f_{\rm peri}$) and an apogalacticton ($f_{\rm apo}$) within 
an integration time of 6 Gyr. Cases that do not have an apogalaticon 
have not completed an orbit, and are considered to be on their first 
infall to the MW. The final two columns list the fraction of orbits 
where two pericenters have occurred ($f_p$) and the average orbital 
period computed using the time of these close passages.

In the previous section, where the LMC was not included, most orbits 
had both an apocenter and pericentric approach to the MW within 6 Gyr. 
Here, we find that the LMC introduces significant scatter to the results. 
In the most extreme case of the lowest MW mass ($M_{\rm MW} = 1.0\times 
10^{12} \Msun$) and the highest LMC mass ($M_{\rm LMC} = 2.5\times 
10^{11} \Msun$), only 9\% of Draco's and 15\% of Sculptor's 10,000 MC 
realizations had closed orbits. In other words, for this light-MW $+$ 
heavy-LMC model, both galaxies were likely on their first approach to 
the MW within the past 6 Gyr. Based on these calculations, we can not 
rule out the possibility that Draco and/or Sculptor are making their 
first approaches to the MW.

Overall, we find that the orbital period and apocenter for both Draco 
and Sculptor systematically increases with the inclusion of the LMC. 
The timing of Sculptor's most recent pericentric approach (0.3--0.4 Gyr ago) 
is a robust quantity, being largely unaffected by changes in MW or LMC mass. 
However, Sculptor's pericentric distance decreases as the LMC mass increases. 
Draco's orbit, on the other hand, is more strongly affected by the 
LMC's inclusion than that of Sculptor. 

\subsection{The Association of Draco and Sculptor with the Disk of Satellites}
\label{ss:dos}

%%%%%%%%%%%%%%%%%%%%%%%%%%%%%%%%%%%%%%%%%%%%%%%%%%%%%%%%%%%%%%%%%%%%%%
%% FIGURE 7
%%%%%%%%%%%%%%%%%%%%%%%%%%%%%%%%%%%%%%%%%%%%%%%%%%%%%%%%%%%%%%%%%%%%%%
%
\begin{figure*}
\subfigure{\includegraphics[scale=0.5]{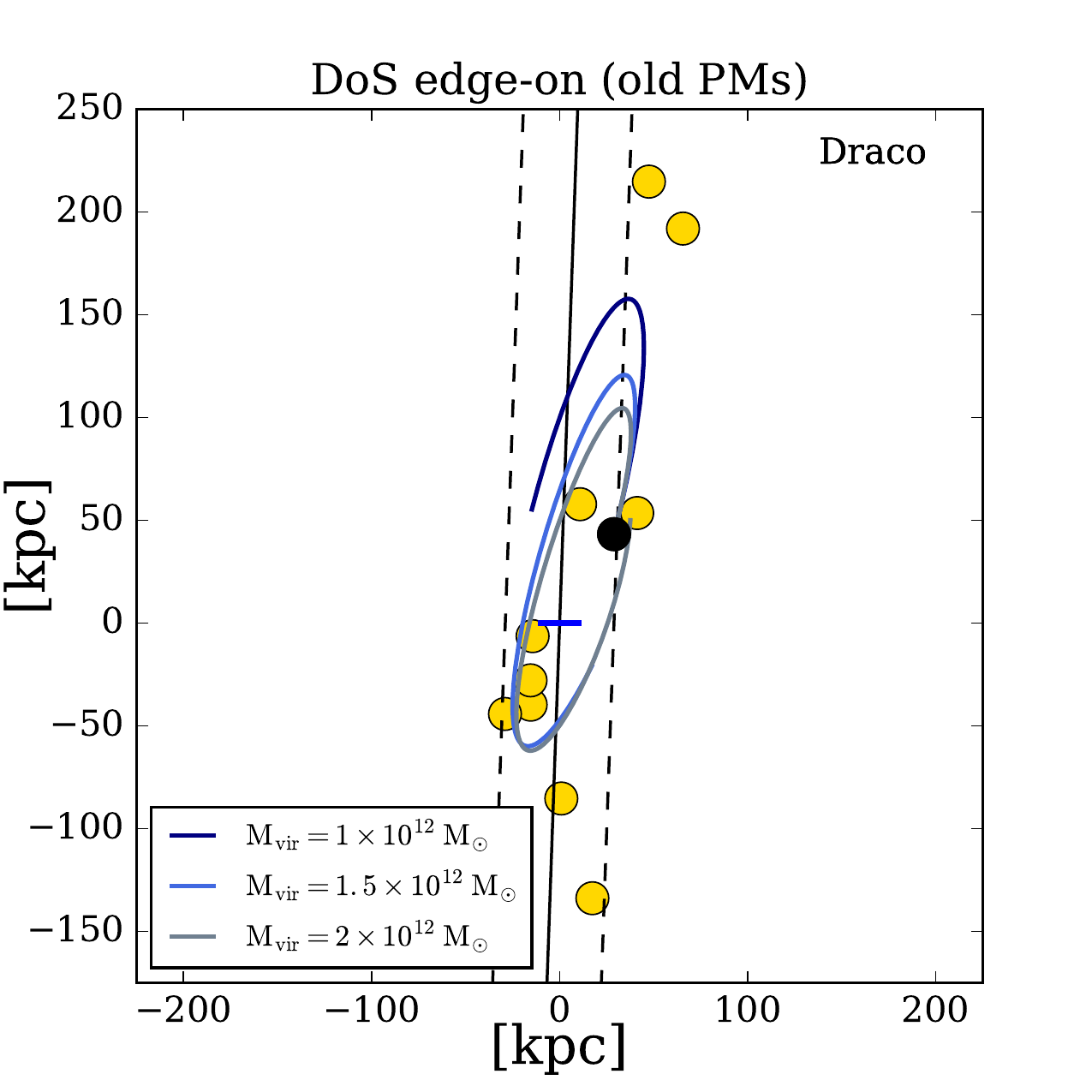}}
\subfigure{\includegraphics[scale=0.5]{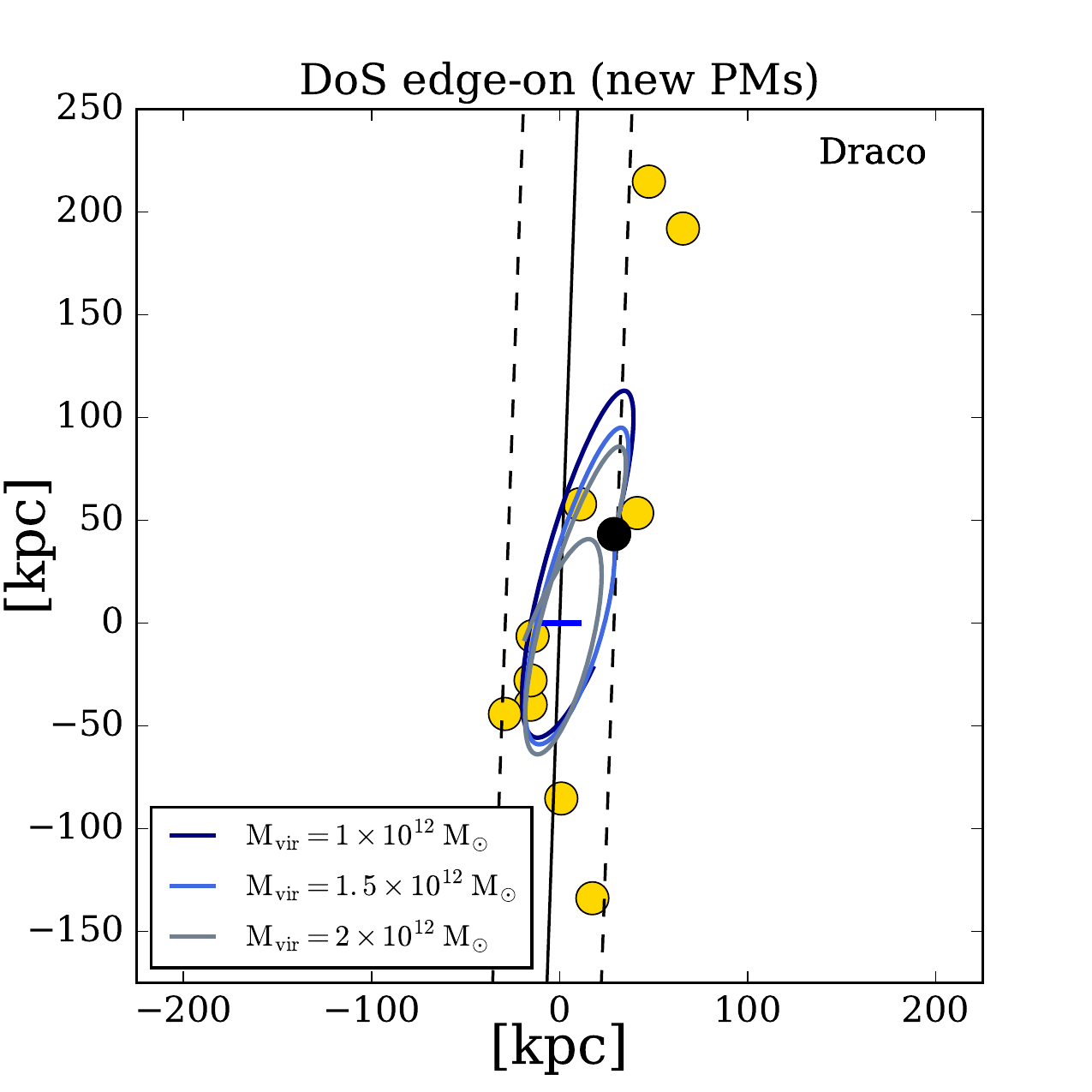}}
\subfigure{\includegraphics[scale=0.5]{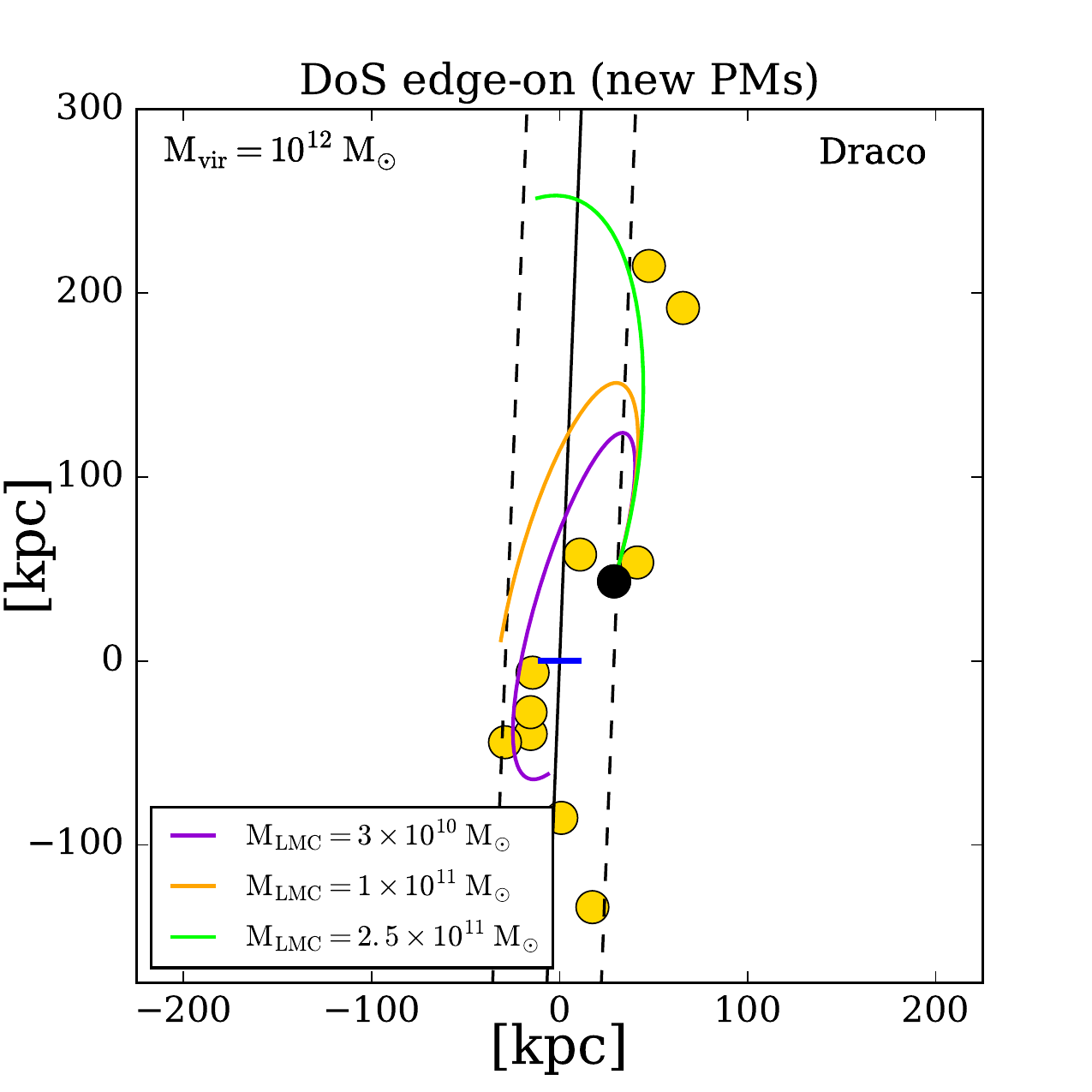}}
\caption{The orbital trajectory of Draco during the last 3 Gyr seen from 
         the edge-on perspective of the DoS (abscissa aligned with $l = 
         156.4\degr$). The nearly vertical lines indicate the best-fitting DoS 
         (solid black lines) and the extents of its height (dashed black lines). 
         The 11 classical MW satellites are indicated with yellow circles. 
         Left panel: Draco's trajectory using the old PM measurement \citet{pry15} 
         in three different MW mass models. Mid panel: Draco's trajectory using 
         the new PM measurement from this work in three MW mass models. 
         Right panel: Draco's trajectory using the new PM and including the 
         gravitational influence of the LMC. In this panel the three trajectories 
         are all calculated in a $1\times10^{12} \Msun$ MW model using three 
         different LMC masses, as listed in the legend. The new PMs result in 
         stronger agreement between the past orbital trajectory of Draco and the 
         DoS over the past 3 Gyr. This statement is robust over the entire mass 
         ranges of the MW and LMC explored in this study.
         \label{f:draco_dos}
}
\end{figure*}
%
%%%%%%%%%%%%%%%%%%%%%%%%%%%%%%%%%%%%%%%%%%%%%%%%%%%%%%%%%%%%%%%%%%%%%%

%%%%%%%%%%%%%%%%%%%%%%%%%%%%%%%%%%%%%%%%%%%%%%%%%%%%%%%%%%%%%%%%%%%%%%
%% FIGURE 8
%%%%%%%%%%%%%%%%%%%%%%%%%%%%%%%%%%%%%%%%%%%%%%%%%%%%%%%%%%%%%%%%%%%%%%
%
\begin{figure*}
\subfigure{\includegraphics[scale=0.5]{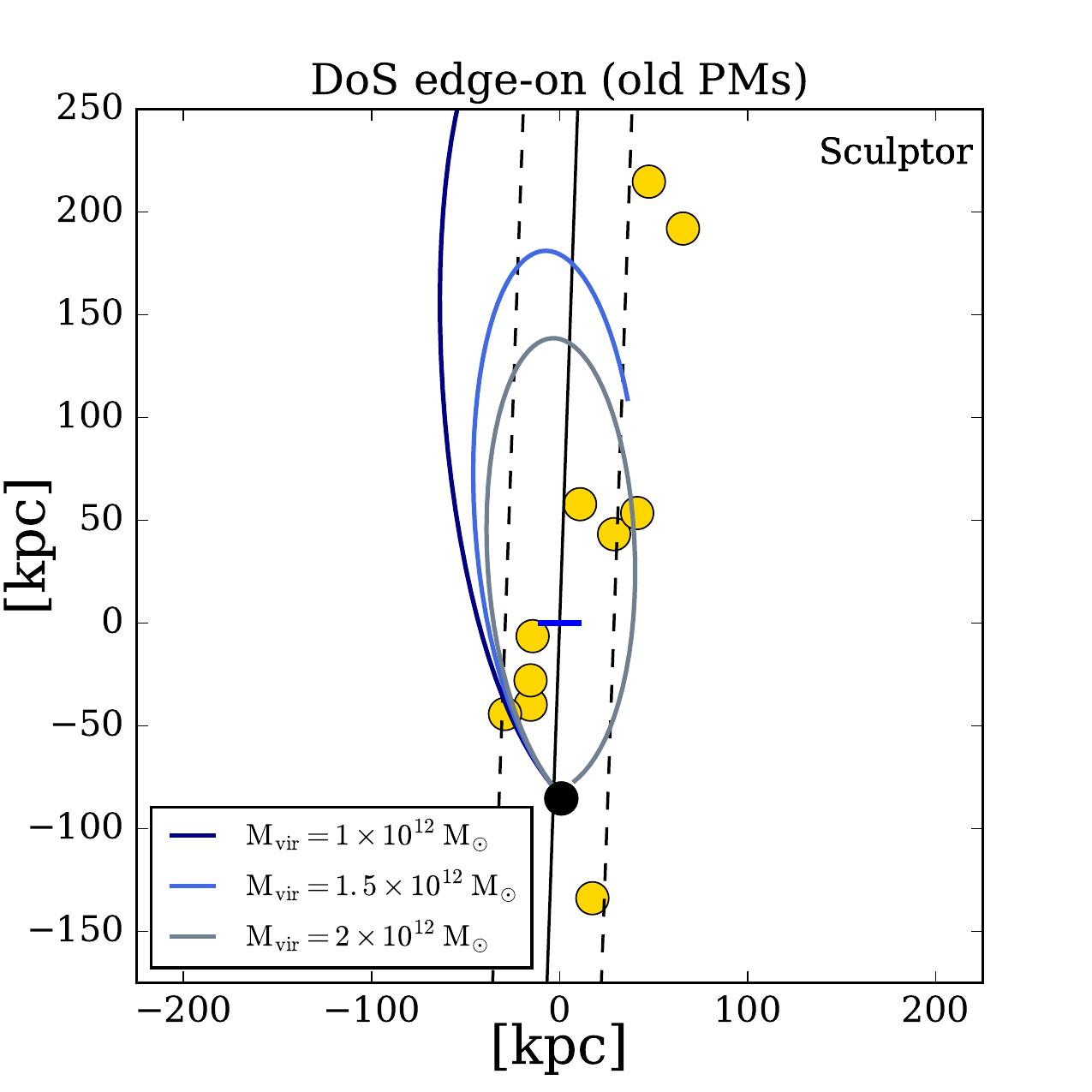}}
\subfigure{\includegraphics[scale=0.5]{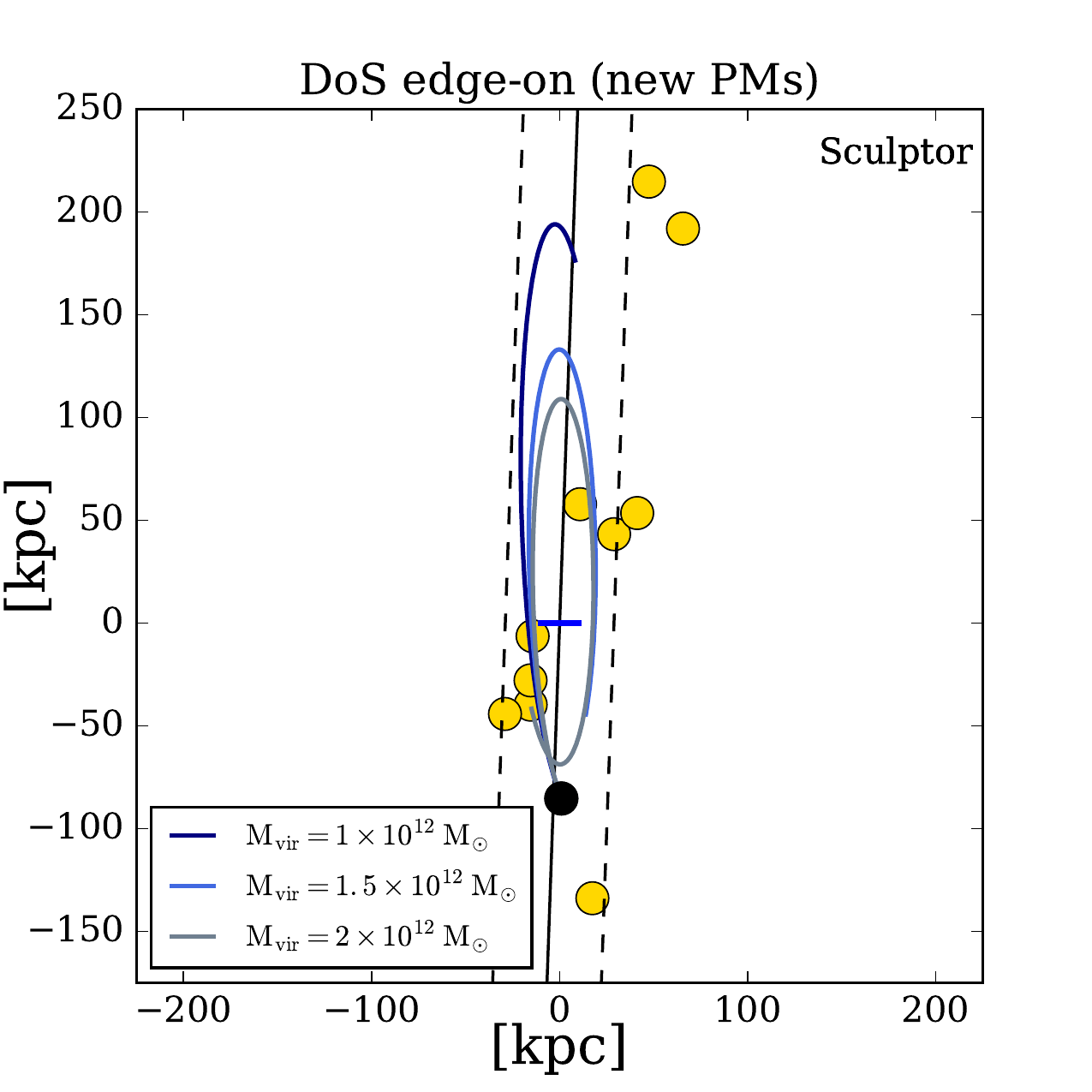}}
\subfigure{\includegraphics[scale=0.5]{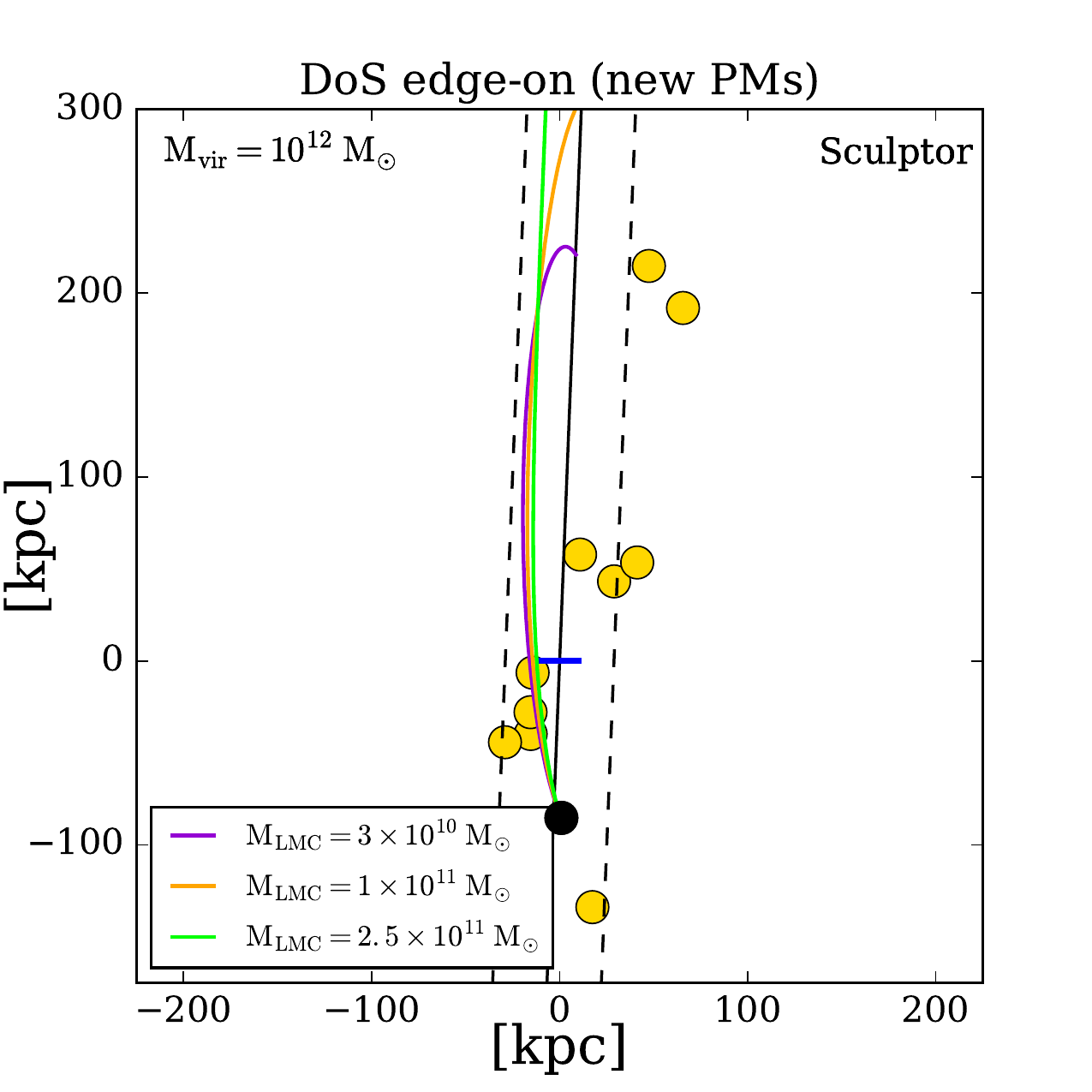}}

\caption{Similar to Figure \ref{f:draco_dos} but now for Sculptor. The old 
         PM measurement for Sculptor comes from \citet{pia06}. As in the case 
         of Draco, the new PMs result in stronger agreement between the past 
         orbital trajectory of Sculptor and the DoS, independent of the MW and 
         LMC masses explored in this study. 
         \label{f:scul_dos}
         }
\end{figure*}
%
%%%%%%%%%%%%%%%%%%%%%%%%%%%%%%%%%%%%%%%%%%%%%%%%%%%%%%%%%%%%%%%%%%%%%%

Draco and Sculptor are classical dSphs that have traditionally been 
included in the DoS. In light of our new PM estimates for these 
satellites, we revisit their dynamical association with the DoS. 
We define the DoS as in \citet{kro10}, where 24 satellite galaxies 
within 254 kpc, including the 11 classical satellites are fit to a 
plane with a minimum disk height of 28.9 kpc. 

Figures~\ref{f:draco_dos} and \ref{f:scul_dos} illustrate the orbital 
trajectory of Draco and Sculptor, respectively, over the past 3 Gyr 
in a viewing perspective such that the DoS is seen edge on. This 
perspective roughly coincides with the Galactocentric X--Z plane.
The current positions of Draco and Sculptor are shown in black dots, 
while the other classical dSphs are shown in yellow. 
We compare orbital trajectories using the previous PM measurements 
(left panels) with those using the new PMs in this study (middle panels).
The previous PM measurements were adopted from \citet{pry15} for Draco, 
and \citet{pia06} for Sculptor, both of which are measured using 
\hst\ data. Orbits are plotted for our 3 different MW mass 
models as indicated in the figure legends. Despite the fact that Draco 
and Sculptor are orbiting in opposite directions about the MW, both 
orbit within the thin DoS for the past 3 Gyr, regardless of the assumed 
MW mass. The agreement between the orbital trajectories of Draco and 
Sculptor and the DoS is substantially improved with the new PMs, 
especially for Sculptor. 

We now examine whether perturbations from the LMC can affect the strong 
agreement between the orbits derived using the new velocity measurements 
and the DoS. In the right panels of Figures \ref{f:draco_dos} and 
\ref{f:scul_dos}, the orbits of Draco and Sculptor for the lowest 
MW mass model ($M_{\rm MW} = 1.0\times 10^{12} \Msun$) are plotted. 
We selected the lowest MW mass model to explore the configuration that 
yields the maximal perturbation on the satellites' orbits by the LMC.
The different trajectories are for the three different LMC mass models 
used in Section~\ref{ss:LMC}. We find that despite increasing the 
LMC mass to as high as $2.5\times10^{11} \Msun$, the orbits of Draco 
and Sculptor are still confined well within the DoS. This indicates that, 
while the LMC can substantially increase the apocenteric distance of the 
orbits of Draco and Sculptor, it does not introduce torques out of their 
orbital plane. This is not surprising since the LMC itself is orbiting 
within the DoS. 

It remains unclear how such a tight agreement between the orbital planes 
of Draco and Sculptor can occur, given that these satellites are orbiting 
in opposite directions about the MW. Our results likely rule out a 
scenario in which Draco and Sculptor were accreted together as part of 
a tightly bound group. Cosmological simulations show that in such a 
scenario, the orbital angular momenta of all group members should be 
well aligned \citep{sal11}. This analysis, however, does not rule out 
that Draco and Sculptor were accreted as part of a loose group or tidal 
structures, which was split apart upon infall \citep[e.g.,][]{paw11}.

Our newly-measured PMs place a new spotlight on an interesting problem. 
While it has been known that Draco and Sculptor are moving in opposite 
directions, we now know that their orbital planes are strongly confined 
within the DoS. Detailed studies of infalling groups of satellites may 
reveal how such satellite orbital configurations are created around 
MW-size galaxies.

\section{Conclusions}
\label{s:conclusions}

We used \hst\ ACS/WFC images to measure the proper motions of 
Draco and Sculptor. By comparing bulk motions of numerous stars in Draco 
and Sculptor with respect to distant background galaxies or QSOs, we 
find the PMs of Draco and Sculptor to be $(\mu_{W},\>\mu_{N})_{\rm Dra} =
  (-0.0562 \pm 0.0099,\>-0.1765 \pm 0.0100)\ {\rm mas\ yr}^{-1}$ and 
$(\mu_{W},\>\mu_{N})_{\rm Scl} =
  (-0.0296 \pm 0.0209,\>-0.1358 \pm 0.0214)\ {\rm mas\ yr}^{-1}$.
These are the most precise PMs measured so far for any satellite dSph  
in the MW halo. We compare our new PM results with previous measurements 
in the literature and find that they are mostly consistent at the 
1--2$\sigma$ levels. However, our results are {\it significant} improvements 
over previous ones with 1d PM uncertainties being at least 5--7 times smaller.

To derive space velocities of Draco and Sculptor in the Galactocentric 
frame, we combined our PMs with known line-of-sight velocities and 
corrected for the solar reflex motions. As a result, our Galactocentric 
radial and tangential velocities are $(V_{\rm rad},\>V_{\rm tan})_{\rm Dra} 
= (-88.6,\>161.4) \pm (4.4,\>5.6) \kms$ and 
$(V_{\rm rad},\>V_{\rm tan})_{\rm Scl} = (72.6,\>200.2) \pm (1.3,\>10.8) \kms$.
We used the total velocities of Draco and Sculptor to provide lower 
limits on the enclosed MW masses at the satellite distances.
The resulting limits are $M > 0.3\times10^{12} \Msun$ and 
$M > 0.5\times10^{12} \Msun$ at distances of $R_{\rm GC} = 76$ kpc 
and 86 kpc, respectively.

We used the PM results to revisit the orbital histories of Draco and 
Sculptor over the past 6 Gyr. Orbital periods of Draco and Sculptor are 
found to be 1--2 and 2--5 Gyrs, respectively, accounting for uncertainties in 
the mass of the MW. The inclusion of the LMC increases the scatter in these 
results. In the most extreme example of a low mass MW ($1.0\times 10^{12} 
\Msun$) and high mass LMC ($2.5\times10^{11} \Msun$), orbital solutions 
favor a scenario where Draco and Sculptor are on their first infall towards 
the MW. The inclusion of the LMC systematically increases the orbital 
period. However, Sculptor's most recent pericentric approach to the MW at 
0.3--0.4 Gyr ago is the most robustly determined orbital property, with 
little variation over a factor of 2 (10) change in halo mass for the 
MW (LMC).

The new PMs measured by this work imply a better agreement between the 
direction of motions of Draco and Sculptor and the purported DoS 
\citep{kro05,met07,kro10,paw13}. Specifically, the new PMs reveal that 
the orbital trajectories of both Draco and Sculptor are confined within 
the DoS for at least the past 3 Gyr. This result is robust to changes in 
MW halo mass and perturbations from the LMC, and likely rule out the 
possibility that Draco and Sculptor were accreted together as part of a 
tightly bound group. 

\acknowledgments

We would like to warmly thank the referee for the constructive 
feedback that helped improve the presentation of our results.
Support for this work was provided by NASA through grants GO-12966
from the Space Telescope Science Institute (STScI), which is operated 
by the Association of Universities for Research in Astronomy (AURA), 
Inc., under NASA contract NAS5-26555. EP is supported by the National 
Science Foundation through the Graduate Research Fellowship Program 
funded by Grant Award No. DGE-1143953.

\facility{HST(ACS/WFC)}.

\end{document}